\documentclass[12pt]{article} 
\pdfoutput=1
\usepackage{epsfig,graphicx,psfrag}
\usepackage{amsmath,amsfonts,stmaryrd,latexsym,amssymb}
\usepackage{url,hyperref,cite,relsize}
\newcommand{\be}{\begin{equation}}
\newcommand{\ee}{\end{equation}}
\newcommand{\bea}{\begin{eqnarray}}
\newcommand{\eea}{\end{eqnarray}}
\newcommand{\bear}{\begin{eqnarray}}
\newcommand{\eear}{\end{eqnarray}}

\newcommand{\ba}{\begin{array}}
\newcommand{\ea}{\end{array}}
\newcommand{\lae}{\begin{array}{c}\,\sim\vspace{-21pt}\\<
\end{array}}

\DeclareMathOperator{\Tr}{Tr}

\begin{document}

\baselineskip=18pt \pagestyle{plain} \setcounter{page}{1}

\vspace*{-1cm}

\noindent \makebox[11.5cm][l]{\small \hspace*{-.2cm}
November 22, 2013; revised June 3, 2014}{\small FERMILAB-Pub-13-350-T} \\ [-2mm]

\begin{center}

{\Large \bf  Higgs mass from compositeness at a multi-TeV scale} \\ [9mm]

{\normalsize \bf Hsin-Chia Cheng$^\diamond$,  Bogdan A. Dobrescu$^\star$, Jiayin Gu$^\diamond$ \\ [4mm]
{\small {\it
$^\diamond$ Department of Physics, University of California, Davis, CA 95616, USA \\ [2mm]
$^\star$ Theoretical Physics Department, Fermilab, Batavia, IL 60510, USA
}}\\
}

\end{center}

\vspace*{0.2cm}

\begin{abstract}
Within composite Higgs models based on the top seesaw mechanism, we show that the  
Higgs field can arise as the pseudo Nambu-Goldstone boson of the broken $U(3)_L$ chiral symmetry 
associated with a vector-like quark and the $t$-$b$ doublet. 
As a result, the lightest CP-even neutral state of the composite scalar sector is lighter than the top quark, and can be 
identified as the newly discovered Higgs boson.
Constraints on weak-isospin violation push the chiral symmetry breaking scale above a few TeV, implying that 
other composite scalars are probably too heavy to be probed at the LHC, but may be within reach at a future hadron collider 
with center-of-mass energy of about 100 TeV. 
\end{abstract}

\vspace*{0.2cm}

{\small \tableofcontents}

\section{Introduction} \setcounter{equation}{0}
\label{sec:intro}

The discovery of the Higgs boson at the Large Hadron Collider (LHC) \cite{Aad:2012tfa} represents an important step towards understanding the origin of electroweak symmetry breaking. Its mass around 126 GeV has major implications for different models or mechanisms of electroweak symmetry breaking. 

A pressing question is whether the Higgs boson is an elementary particle or a composite one. 
Current data cannot distinguish between an elementary Higgs boson (up to very high energies) and a composite Higgs boson arising from 
a theory with a decoupling limit identical to the Standard Model (SM). Nonetheless, the requirement that the Higgs boson mass $M_h$ 
is near 126 GeV poses model building challenges on SM extensions.

A composite Higgs boson has generically a large quartic coupling and hence is expected to be heavy, unless its mass is protected by some symmetry. 
One possibility is that the Higgs field arises as a pseudo-Nambu-Goldstone boson (pNGB) of some broken global symmetry
\cite{Kaplan:1983fs}.  Many types of models which realize the Higgs multiplet as a pNGB have been proposed,
including Little Higgs \cite{ArkaniHamed:2001nc} 
and models that invoke the AdS/CFT correspondence \cite{Contino:2003ve}. 
In these effective theories, $M_h$ is quite model-dependent. 
The quartic coupling arises at tree level in some models, and from loops in others. The relatively large top quark mass, $m_t$, can be accommodated through partial compositeness \cite{Kaplan:1991dc}, {\it i.e.},  mixing of the 
elementary top fields with composite vector-like fermions. In a class of pNGB composite Higgs models where the dominant explicit breaking of the global symmetry comes from these mixings \cite{Contino:2003ve},  
there are strong correlations between $M_h$ and the masses of composite vector-like fermions. Typically, $M_h \approx 126$ GeV requires at least one vector-like top partner well below 1 TeV \cite{Contino:2006qr} which is in some tension with 
the 8 TeV LHC data, and will be thoroughly tested by LHC searches in the near future. An additional prediction of these models that will 
be soon tested is that higher dimensional operators modify the Higgs couplings \cite{Giudice:2007fh}.

A more direct way of obtaining a heavy top quark is the top condensation mechanism, where the Higgs field is a top-antitop
bound state \cite{Nambu:1989jt, Bardeen:1989ds}.
The top quark, however, would have in this case a mass around 600 GeV if the compositeness scale is not much above the electroweak scale. Furthermore, the leading number of colors ($N_c$) approximation gives $M_h \approx 2 m_t$. Both results are in conflict with the data. The top and Higgs masses can be reduced, at the expense of fine-tuning, if the compositeness scale is raised; but even with compositeness at the Planck scale, $m_t$ and $M_h$ would still be above 200 GeV \cite{Bardeen:1989ds}. 

An attractive solution to the top mass problem in top-condensation models is through the top-seesaw mechanism 
\cite{Dobrescu:1997nm, Chivukula:1998wd, Dobrescu:1999gv, He:2001fz}. The heavy top given by top condensation mixes with a massive vector-like top partner, resulting in a light mass eigenstate of $m_t \approx  173$~GeV, identified as the observed top quark. Although the composite Higgs boson is likely to be heavier than the weak scale, for some range of parameters it is possible to obtain a relatively light Higgs boson
\cite{Chivukula:1998wd, Dobrescu:1999gv, Fukano:2013kia}. 
Generically, the composite Higgs sector includes two doublets and two singlets \cite{Dobrescu:1999gv}, and a light Higgs boson could arise 
due to the mixing among the four CP-even scalars. 

In this paper we show that a composite Higgs boson can naturally have $M_h \approx 126 $ GeV (once the weak scale $v$ is fixed at 246 GeV)  
in a top-seesaw model where the strong dynamics 
preserves a global  $U(3)_L$ symmetry acting on the top-bottom doublet $(t_L, \, b_L)$ and the left-handed component of the vector-like quark, $\chi_L$. The composite Higgs doublet is the collection of pNGB's of the spontaneous $U(3)_L \to U(2)_L$ breaking. The explicit symmetry breaking required for the seesaw mechanism also controls $M_h$. As a consequence, the mass of the composite Higgs boson is correlated with $m_t$, and has an upper bound around $m_t$. The electroweak interactions further reduce $M_h$ so that generically it is substantially below $m_t$. 

The low energy effective theory is given by the SM plus the heavy states arising from two Higgs doublets and two complex singlet scalars, together with the  vector-like top partner. The strongest constraint on this model is due to weak-isospin violation, which requires 
the vector-like quark and consequently most of the scalars other than the SM Higgs boson to have masses of order 10 TeV. This implies some tuning to obtain the weak scale at $v \approx 246$ GeV. Nevertheless, given that no new physics has been found at the LHC so far, some tuning seems inevitable in any theory which attempts to explain electroweak symmetry breaking.

The strong dynamics responsible for compositeness and chiral symmetry breaking may be an asymptotically-free gauge interaction that is spontaneously 
broken near the scale $\Lambda$ where it becomes strong. Various theories of these type have been proposed \cite{Hill:1991at,Hill:2002ap}.
It may even be possible that a strongly coupled gauge interaction breaks 
its own gauge symmetry \cite{Martin:1992aq, ArkaniHamed:2000hv}. The consequence would be a non-confining strong interaction whose 
bound states are present together with their constituents at energies below the compositeness scale $\Lambda$. 
Its effects at the scale $\Lambda$ may be parametrized by certain 4-fermion interactions.
Such a theory is viable because in the decoupling limit, where the scale of chiral symmetry breaking $f$ is much larger than $v$,
the deviations from the SM are  $O(v^2/f^2)$. In practice, the hierarchy of scales $v < f < \Lambda$ should not be too large 
because both $f/\Lambda$ and $v/f$ require fine-tuning.

In Section \ref{sec:chs} we present the effective theory of the composite Higgs model below the compositeness scale. We discuss the $U(3)_L \times U(2)_R$ symmetric terms and also their explicit breaking terms. In Section \ref{sec:ewsb} we find the vacuum which breaks the electroweak symmetry and produces the top-seesaw mechanism. In Section \ref{sec:MH} we derive an approximate analytic formula for $M_h$, and find that  its maximum value is near $m_t$. In Section \ref{sec:EW} we discuss the effects of electroweak gauge loops and show that they always reduce $M_h$.
In Section \ref{sec:spectrum} we perform a detailed numerical study of $M_h$ as a function of the parameters in this model. We show that the 126 GeV Higgs boson can be easily accommodated as a composite field in our model. After fixing $M_h \approx 126$ GeV, we calculate the spectrum of the heavy states, 
and then in Section 5 we comment on phenomenological implications. The conclusions are drawn in Section \ref{sec:conclusions}.

\section{Composite Higgs from $U(3)_L\times U(2)_R$ symmetric dynamics}\setcounter{equation}{0}
\label{sec:chs}

We consider an effective theory at a scale $\Lambda \gg 1 $ TeV that includes the SM gauge group and fermions, 
an $SU(2)_W$-singlet vector-like quark, $\chi$, of electric charge $+2/3$, and some 4-fermion interactions
suppressed by $\Lambda$ (which presumably arise due to an asymptotically-free gauge interaction that is spontaneously 
broken near the scale $\Lambda$ where it becomes strong). This theory does not include a Higgs doublet nor any elementary scalars.
We assume that some of the 4-fermion interactions involving third generation quarks and $\chi$ 
are attractive and sufficiently strong to form quark-antiquark  bound states  \cite{Dobrescu:1999gv}.
These strong interactions are not confining, because at distances longer than $1/\Lambda$ the effects
of 4-fermion interactions (other than the presence of bound states) are exponentially suppressed. 

The low-energy theory includes the composite fields that are  deeply bound such that their masses are 
less than the compositeness scale $\Lambda$. As in the Nambu-Jona-Lasinio (NJL) model \cite{Nambu:1961tp},
if the coefficient of a 4-fermion interaction is larger than a certain critical value, then the squared mass of the scalar bound state 
becomes negative, and the scalar acquires a vacuum expectation value (VEV). Furthermore, as the coefficient crosses the critical value, 
there is  a second order phase transition so that the VEV can in principle be much smaller than $\Lambda$ \cite{Hill:2002ap}.

Concretely, we assume the constituents of the deeply bound states to be only 
$\chi_L$, $\chi_R$, the right-handed top quark $t_R$, and the left-handed top-bottom doublet  $\psi^3_L = (t_L, b_L)$.
In the limit where the electroweak interactions are ignored, the kinetic terms of these quarks have an $U(3)_L \times U(2)_R$ 
chiral symmetry, which we assume to be approximately preserved by the 4-fermion interactions. 
The $U(3)_L \times U(2)_R$ symmetric interactions give rise to the following Yukawa couplings of the composite scalars (collectively labelled by $\Phi$) to their constituents:
\begin{equation}
{\cal L}_{\rm Yukawa} = - \xi \left( \bar{\psi}^3_L , \bar{\chi}_L \right)  \Phi \begin{pmatrix} t_R \\ \chi_R \end{pmatrix} + {\rm H.c.}    
\label{eq:Vtop}
\end{equation}
Here $\xi$ is a dimensionless coupling whose value at scale $\Lambda$, upon integrating out  $\Phi$, matches the coefficient of the 4-fermion interactions.
The scalar field $\Phi$ is a $3\times 2$ complex matrix
\begin{equation}
\Phi = \left( \Phi_t  \;\, , \; \Phi_\chi  \right)  ~~,
\end{equation}
where the scalar fields $\Phi_t$ and $\Phi_\chi$ are the bound states of the $U(3)_L$ triplet $(t_L, b_L, \chi_L)$ with $t_R$ and $\chi_R$, respectively:
\begin{equation}
\Phi_t \sim \bar{t}_R \begin{pmatrix} \psi^3_L \\ \chi_L \end{pmatrix}  \hspace{1cm},\hspace{1cm}
\Phi_\chi \sim \bar{\chi}_R \begin{pmatrix} \psi^3_L \\ \chi_L \end{pmatrix} ~~.
\end{equation}

At scales $\mu < \Lambda$, the Yukawa couplings (\ref{eq:Vtop}) give rise to the following  potential for $\Phi$:
\begin{equation}
V_\Phi =  \frac{\lambda_1}{2} \Tr \!\left[ (\Phi^\dagger \Phi)^2 \right] + \frac{\lambda_2}{2} \left({\rm Tr}[\Phi^\dagger \Phi]  \right)^2  + M_\Phi^2 \Phi^\dagger \Phi ~~.
 \label{eq:quartic}
\end{equation}
The quartic couplings $\lambda_1$ and $\lambda_2$ depend on the scale $\mu$; if the kinetic term for $\Phi$ is canonically normalized, then 
$\lambda_1$  becomes non-perturbative near $\Lambda$.  In the large $N_c$ limit, $\lambda_1$ 
is generated by a fermion loop, while $\lambda_2$ vanishes. Scalar loops, however, generate a non-zero value for $\lambda_2$,
so that $\lambda_1 \gg |\lambda_2|$.
 In the Appendix 
 we use 1-loop renormalization group (RG) equations to estimate the $\lambda_1/\lambda_2$ ratio.  

The squared mass of $\Phi$ is assumed to satisfy $|M_\Phi^2| \ll \Lambda^2$,
and if the 4-fermion interactions are super-critical, then $M_\Phi^2 < 0$ triggering spontaneous symmetry breaking of $U(3)_L \times U(2)_R$.

We assume that there are additional explicit $U(2)_R$ breaking effects which distinguish $t_R$ and $\chi_R$.
Given that $|M_\Phi | \ll \Lambda$, 
such effects could induce a large relative splitting of the masses for  $\Phi_t$  and $\Phi_\chi$. We parametrize these effects by 
\begin{equation}
V_{ {\rule[2pt]{16pt}{0.5pt}} \hspace{-16pt} U(2) } = \delta M^2_{tt}  \; \Phi^\dagger_t \Phi_t + \delta M^2_{\chi\chi} \; \Phi^\dagger_\chi \Phi_\chi  + (M_{\chi t}^2 \Phi^\dagger_{\chi} \Phi_t + {\rm H.c.} )
\label{eq:mass}
\end{equation}
These $U(2)_R$ breaking masses can be diagonalized by a $U(2)$ rotation. As we will see later, it is convenient to work in a different basis from the one 
that diagonalizes these mass terms so we will keep Eq.~(\ref{eq:mass}) general.

The $U(2)_R$ breaking effects may also split the quartic couplings in (\ref{eq:quartic}), but this does not make any qualitative difference 
(unless the changes in couplings are larger than order one), because our discussion mostly relies on the $U(3)_L$ symmetry. For simplicity we do not include such effects.

Gauge invariant masses for the $SU(2)_W$-singlet quarks can be present at the scale $\Lambda$:
\begin{equation}
{\cal L}_{\rm mass} = -\mu_{\chi t} \bar{\chi}_L t_R - \mu_{\chi\chi} \bar{\chi}_L \chi_R + {\rm H.c.}
\label{eq:fermion_mass}
\end{equation}
We assume that $\mu_{\chi t}, \mu_{\chi \chi} \ll \Lambda$ (which is technically natural because there is an enhanced chiral symmetry in the 
$\mu_{\chi t}, \mu_{\chi \chi}  \to 0$ limit), so that the tree-level quark masses do not disrupt the formation of bound states. 
The above mass terms break $U(3)_L \times U(2)_R$ down to $U(2)_L \times U(1)_R$.
Below $\Lambda$, these fermion masses map to tadpole terms for the $SU(2)_W$-singlet scalars: 
\begin{equation}
V_{\rm tadpole} =  - (0,0,C_{\chi t})\Phi_t -  (0,0,C_{\chi\chi})\Phi_\chi + {\rm H.c.} 
\end{equation}
Matching at the scale $\Lambda$, we have 
\be
C_{\chi t} \simeq  \frac{\mu_{\chi t}}{\xi} \Lambda^2 \;\;\; , \;\;\; C_{\chi \chi} \simeq  \frac{\mu_{\chi \chi}}{\xi} \Lambda^2 \, .
\label{eq:tadpole_matching}
\ee
Note that when the scalars are  integrated out at the cutoff scale (where $M_{\Phi}^2 \sim \Lambda^2$), the fermion mass terms (\ref{eq:fermion_mass})
are recovered.

The effective potential of the scalar sector below the compositeness scale is given by
\begin{eqnarray}
V_{\rm scalar} &=& \frac{\lambda_1+ \lambda_2}{2} \left[ (\Phi^\dagger_t \Phi_t)^2+(\Phi^\dagger_\chi \Phi_\chi)^2 \right]
+ \lambda_1 |\Phi^\dagger_t \Phi_\chi|^2  + \lambda_2 (\Phi^\dagger_t \Phi_t)(\Phi^\dagger_\chi \Phi_\chi) \nonumber \\ [2mm]
&& + \, M^2_{tt} \Phi^\dagger_t \Phi_t + M^2_{\chi\chi} \Phi^\dagger_\chi \Phi_\chi  + (M_{\chi t}^2 \Phi^\dagger_{\chi} \Phi_t + {\rm H.c.}) \nonumber \\ [2mm]
&& - \, (0,0, 2C_{\chi t}) \,  {\rm Re}\, \Phi_t  - (0,0, 2C_{\chi \chi}) \,  {\rm Re}\, \Phi_\chi    ~~,
\label{eq:V}
\end{eqnarray}
where $M^2_{tt}$ and $M^2_{\chi\chi}$ are the sums of the mass terms in Eqs.~(\ref{eq:quartic}) and (\ref{eq:mass}). 
The parameters $M^2_{\chi t}$, $C_{\chi t}$, and $C_{\chi\chi}$ can be chosen to be real without loss of generality. As mentioned earlier, $M^2_{\chi t}$ can be removed by a $U(2)$ rotation. Then $C_{\chi t}$ and $C_{\chi\chi}$ can be made real by phase redefinitions of $\Phi_t$ and $\Phi_\chi$ respectively. The effective field theory below the cutoff $\Lambda$ is thus described by Eqs.~(\ref{eq:Vtop}) and (\ref{eq:V}).

The $SU(2)_W \times U(1)_Y$ gauge symmetry is a subgroup of the $U(3)_L \times U(2)_R$ chiral symmetry. Thus, the electroweak interactions explicitly break
the $U(3)_L$ symmetry. The $U(3)_L$ triplets $\Phi_t$ and $\Phi_\chi$ can be written in terms of fields belonging to electroweak representations:
\begin{equation}
\Phi_t = \begin{pmatrix} H_t \\ \phi_t \end{pmatrix}   \hspace{1cm}, \hspace{1cm}
\Phi_\chi = \begin{pmatrix} H_\chi \\ \phi_\chi \end{pmatrix}  ~~.
\end{equation}
$H_t$ and $H_\chi$ transform under $SU(2)_W \times U(1)_Y$ as the SM Higgs doublet, while $\phi_t$ and $\phi_\chi$ are $SU(2)_W \times U(1)_Y$ singlets. 
The effects of the electroweak interactions will be discussed in Sec.~\ref{sec:EW}.
$H_{\chi (t)}, \, \phi_{\chi (t)}$ will get VEVs due to negative squared masses and tadpole terms. Expanding $H_{\chi (t)}, \, \phi_{\chi (t)}$ around their VEVs in terms of fields of definite electric charges, we can write  
\begin{eqnarray}
H_t =  \begin{pmatrix} \displaystyle  \frac{1}{\sqrt{2}} \left(v_t + h_t +iA_t \right) \\ H^-_t \end{pmatrix}   \hspace{1cm} , && \hspace{1cm}
H_\chi =  \begin{pmatrix} \displaystyle  \frac{1}{\sqrt{2}} \left(v_\chi + h_\chi +iA_\chi \right) \\ H^-_\chi \end{pmatrix}  ~~,  
 \nonumber \\ [3mm]
\phi_t = \frac{1}{\sqrt{2}}(u_t + \varphi_t +i \pi_t)  \hspace{0.8cm} \hspace{1cm} , && \hspace{1.1cm}
\phi_\chi = \frac{1}{\sqrt{2}}(u_\chi + \varphi_\chi +i \pi_\chi)   ~~.
\label{eq:linear}
\end{eqnarray}
The VEVs $v_t, v_\chi, u_t$ and $u_\chi$ are real, and some of them may vanish, depending on the parameters of the effective potential.
We use the notation $v_t^2 + v_\chi^2 = v^2$, $u_t^2 + u_\chi^2 = u^2$, and 
\be
f = \sqrt{u^2 + v^2}   
\label{eq:f}
\ee
is the scale of $U(3)_L$ breaking. The measured Fermi constant requires $v \approx 246$ GeV.

We now analyze the low energy effective theory given by Eqs.~({\ref{eq:V}) and ({\ref{eq:Vtop}).

\subsection{Electroweak symmetry breaking}
\label{sec:ewsb} 

It is convenient to perform an $U(2)_R$ transformation (which rotates $t_R$ and $\chi_R$, as well as $\Phi_t$ and $\Phi_\chi$) to go to a basis where $v_t=0$ and $v_\chi=v$. 
For simplicity, we  will use the same notation in this basis as in Eqs.~(\ref{eq:V})-(\ref{eq:linear}). In this basis we define 
$u_t = u \sin \gamma$ and $u_\chi = u \cos \gamma$, and the short-hand notation $s_\gamma= \sin\gamma$ and $c_\gamma= \cos\gamma$. 

The extremization conditions for $V_{\rm scalar}$ relate the parameters from the effective potential to the VEVs:
\bear
&& v \left( M_{\chi t}^2 + \frac{\lambda_1}{2}  u^2 s_\gamma c_\gamma \right) =0 ~ \, ,
\nonumber \\ [2mm]
&& v \left( M_{\chi \chi}^2 +\frac{\lambda_1}{2}  \left( u^2 c_\gamma^2 + v^2 \right) + \frac{\lambda_2}{2}  \left( u^2+v^2 \right)\right)=0\, ~ ,
\label{eq:tadpole_eq2}
\eear
for the derivatives with respect to $h_t$ and $h_\chi$, and 
\bear
&& C_{\chi t} =  \frac{u}{\sqrt{2} } \left[ M^2_{\chi t} \, c_\gamma + \left( M_{tt}^2 + \frac{\lambda_1}{2} u^2 + \frac{\lambda_2}{2}  \left(u^2 +v^2 \right)\right) s_\gamma \right]  \, ~,
\nonumber \\ [2mm]
&& C_{\chi \chi} =  \frac{u}{\sqrt{2} } \left[ M_{\chi t}^2 \, s_\gamma + \left( M_{\chi \chi}^2 + \frac{\lambda_1 +\lambda_2}{2}  \left( u^2+v^2 \right) \right) c_\gamma \right] \, ~,
\label{eq:tadpole_eq4}
\eear
for the derivatives with respect to $\varphi_t$ and $\varphi_\chi$.
Eqs.~(\ref{eq:tadpole_eq2}) and (\ref{eq:tadpole_eq4}) have a solution for $v=0$, and a different solution for $v > 0$. When the latter is 
a minimum ({\it i.e.}, the squared masses of all spin-0 states are positive), we find that it is also 
the global minimum of the effective potential while $v=0$ is a saddle point. 

For $v > 0$, Eqs.~(\ref{eq:tadpole_eq2})  imply
\bear
M_{\chi t}^2 &=& - \frac{ \lambda_1}{2} u^2 s_\gamma c_\gamma \, ~,
\nonumber \\ [2mm]
M_{\chi \chi}^2 &= &-\frac{ \lambda_1}{2} \left(u^2 c_\gamma^2 + v^2\right) - \frac{ \lambda_2}{2}  \left(u^2+v^2\right) \, ~.
\label{eq:sol_Mxx}
\eear
Substituting these into Eqs.~(\ref{eq:tadpole_eq4}) gives
\bear
C_{\chi t} &=&  \frac{u \, s_\gamma}{\sqrt{2} } \left[  M_{tt}^2 +  \frac{\lambda_1}{2} u^2 s_\gamma^2 + \frac{\lambda_2}{2} \left(u^2+v^2\right) \right] \, ~,
\nonumber \\ [2mm]
C_{\chi\chi} &=& 0 \, ~.
\label{eq:sol_Cxx}
\eear
Thus, the basis where $\langle H_t \rangle = 0$ and $\langle H_\chi \rangle \neq 0$
is the one where $C_{\chi\chi}=0$
(or equivalently, where $t_R$ and $\chi_R$ are defined such that $\mu_{\chi\chi} = 0$).
Since the electroweak symmetry is broken 
only by the VEV of $H_\chi$, the eaten Nambu-Goldstone bosons are contained in $H_\chi$ only. 
The charged Higgs boson resides entirely within $H_t$, and its mass squared is
\be
M_{H^\pm}^2 = M_{tt}^2 +\frac{ \lambda_1}{2}  u^2  s_\gamma^2 + \frac{ \lambda_2}{2} \left(u^2+v^2\right) \, .
\label{eq:charged_Higgs_mass}
\ee
The nonzero tadpole coefficient is then related to $M_{H^\pm}^2$ by  $ C_{\chi t} =  u \, s_\gamma  M_{H^\pm}^2 / \sqrt{2} $.
Note that $v \neq 0$ requires $M_{\chi\chi}^2 < 0$, but does not restrict the sign of $M_{tt}^2$.
In Section~\ref{sec:spectrum}, however, we will show that the Higgs boson would be lighter than about 100 GeV unless 
$M_{H^\pm} > u$, which in turn  requires $M_{tt}^2 > 0 $.

The constraint from weak-isospin violation (discussed in Section~\ref{sec:spectrum}) requires $v \ll u$. As a result, 
the $U(3)_L$ breaking scale, defined in Eq.~(\ref{eq:f}), is given by $f \simeq u$.
Eqs.~(\ref{eq:sol_Mxx}) and (\ref{eq:sol_Cxx}) have solutions for $v$, $u$ and $s_\gamma$ that satisfy 
$v/u \ll 1$ provided there is a fine-tuned relation among the  $M_{\chi \chi}$, $M_{\chi t}$,   $M_{tt}$ and 
$C_{\chi t}$ parameters. 

Neglecting the mixing of the charm and up quarks with $t$ and $\chi$, the mass terms of the heavy
charge-2/3 quarks, arising from Eq.~(\ref{eq:Vtop}), are given by
\be
- \frac{\xi}{\sqrt{2}}  \left( \overline{t}_L , \overline{\chi}_L \right)  \begin{pmatrix} 0 & v  \\  u s_\gamma  & u c_\gamma \end{pmatrix}   \begin{pmatrix} t_R  \\  \chi_R \end{pmatrix}  + {\rm H.c.}  
\ee
Diagonalizing this matrix gives the masses of the top quark $t$ and the new quark, which we label by $t^\prime$ in the mass eigenstate basis. 
Keeping only the leading  nonvanishing term in $v^2/f^2$, we find 
the mass of the top quark, 
\be
m_t \simeq  \frac{\xi  }{\sqrt{2}} \, v s_\gamma   ~.
\ee
Thus, $\xi$ and $s_\gamma$ can be related to the top Yukawa coupling $y_t$ by
\be
s_\gamma \simeq \frac{y_t}{\xi} ~.
\label{sgamma}
\ee
$\xi$ is expected to have a value around 3 or 4 (see Section~\ref{sec:spectrum}) and $y_t \sim 1$, so  $s_\gamma^2 \sim  O(0.1)$. 
Again to leading order in $v^2/f^2$, 
the mass of the new quark is given by
\be
m_{t^\prime} \simeq \frac{\xi }{\sqrt{2}} \,  f   ~.
 \label{eq:mt'}
\ee
while the mixing angle $\theta_L$, which rotates the $t_L$ and $\chi_L$ gauge eigenstates into the mass eigenstate quarks, 
is given by
\be
\sin\theta_L \equiv s_L \simeq \frac{v}{f} ~.  \label{eq:thetaL}
\ee

\subsection{Analytical expression for the Higgs mass}
\label{sec:MH}

Substituting Eqs.~(\ref{eq:sol_Mxx})-(\ref{eq:charged_Higgs_mass})  back into the scalar potential~(\ref{eq:V}), 
we find that the $4\times 4$ mass-squared matrix of the CP-even neutral scalars $(h_t, \, h_\chi, \, \varphi_t, \, \varphi_\chi)$ is given by
\be
\left( \ba{cccc}  \displaystyle
\!\! M_{H^\pm}^2 + \! \frac{ \lambda_1}{2} v^2 & 0 &  \displaystyle -\frac{\lambda_1}{2}  u v c_\gamma &  \displaystyle - \frac{\lambda_1}{2} u v s_\gamma \\  [3mm]
0 & (\lambda_1+\lambda_2) v^2 & \lambda_2 u v s_\gamma & (\lambda_1+\lambda_2) u v c_\gamma \\ [3mm]
 \displaystyle  -\frac{\lambda_1}{2}  u v  c_\gamma & \lambda_2 u v  s_\gamma 
 &  \displaystyle M_{H^\pm}^2 \!\! +\! \left[\lambda_1\!\!\left(\! 1\!-\! \frac{c_\gamma^2}{2}\right) \!+\! \lambda_2 s_\gamma^2 \right] \! u^2 
 &  \displaystyle \left( \frac{\lambda_1}{2} + \lambda_2 \right) u^2 s_\gamma c_\gamma \\  [4mm]
 \displaystyle - \frac{\lambda_1}{2}  u v s_\gamma & (\lambda_1+\lambda_2) u v c_\gamma 
 &  \displaystyle \left( \frac{\lambda_1}{2} + \lambda_2\right)  u^2 s_\gamma c_\gamma 
 &  \displaystyle
 \left[\lambda_1 \!\!\left(\!1\!-\! \frac{s_\gamma^2}{2}\right) \!  + \lambda_2 c_\gamma^2 \right] \! u^2  \!\!
\ea \right) ~.
\label{eq:cpeven_mass_matrix}
\ee

The lightest Higgs boson mass-squared ($M_h^2$) is given by the smallest eigenvalue of the mass matrix~(\ref{eq:cpeven_mass_matrix}). 
Using Eq.~(\ref{sgamma}), keeping the leading order in $v^2/f^2$,  and expanding in $s_\gamma$, we find
\be
M_h^2 =  \frac{ \lambda_1 v^2 s_\gamma^2  \, M_{H^\pm}^2}{2 M_{H^\pm}^2 \!\! + \! \lambda_1 u^2 } \,
\!\!
\left[ 1+  \frac{(\lambda_1+2\lambda_2) (M_{H^\pm}^2 \!\! + \lambda_1 u^2 )^2 }
{(\lambda_1+\lambda_2)M_{H^\pm}^2 \, ( 2 M_{H^\pm}^2 \!\! + \!\lambda_1  u^2 ) }  s_\gamma^2+ {\cal O} (s_\gamma^4)  \right]  ~~.
\label{eq:Higgsmass}
\ee  
The Higgs boson mass is suppressed by $s_\gamma$, because in the limit of $\xi\to \infty$ or $m_t \to 0$ the explicit $U(3)_L$ breaking tadpole terms $C_{\chi t}$, $C_{\chi\chi}$ vanish and the $H_\chi$ and $\pi_\chi$ fields become Nambu-Goldstone bosons. Keeping only the leading term in $s_\gamma^2$,
\bear
M_h^2 & \simeq & \frac{\lambda_1}{2 \xi^2}   \left(1 + \frac{\lambda_1  m_{t'}^2 }{\xi^2  M_{H^\pm}^2 } \right)^{\! -1}  y_t^2 \, v^2
\nonumber \\
& = &  \lambda_h \, v^2 ~~ ,
\label{eq:light_Higgs_mass}
\eear
where $\lambda_h$ is the Higgs quartic coupling. In the fermion-loop approximation of NJL,  
the ratio of couplings $\lambda_1/ (2\xi^2)$ is equal to 1. Scalar and gauge boson loops reduce this ratio. In the Appendix
we show that it has a quasi-infrared fixed point value $\sim 0.4$, so we expect $\lambda_1/ (2\xi^2)$ between 0.4 and 1.

Since both $y_t$ and $\lambda_h$  are obtained from integrating out the heavy quark, Eq.~(\ref{eq:light_Higgs_mass}) relates the quartic Higgs coupling  and the top Yukawa coupling at the scale $m_{t'}$, implying $\lambda_h < y_t^2$ at that scale. For $m_{t'}\sim 10$~TeV, we find $y_t^2 \sim 0.6$. Evolving $\lambda_h$ down to the scale $v$ we obtain an upper limit on the Higgs mass,
\be
M_h \lae 185 \mbox{ GeV}  ~~.  
\label{eq:mhmax}
\ee
This is an interesting result, in contrast to the na\"ive expectation in many composite Higgs models that the Higgs boson is heavier than the weak scale. We see that the Higgs boson is light because it is a pNGB of the $U(3)_L\to U(2)_L$ symmetry breaking. In Section 4 we perform a more refined analysis, and as a result
the above upper limit is further reduced.

There is one additional pNGB, $A_1$, mostly given by the $\pi_\chi$ field, with small admixtures of the other neutral CP-odd scalars in Eq.~(\ref{eq:linear}).
Its squared mass is 
\be
M^2_{A_1} = \frac{1}{4}\left( 2 M^2_{H^\pm} + \lambda_1 f^2 \right) \left[ 1
- \sqrt{ 1 - \frac{8 \lambda_1 M^2_{H^\pm} f^2 s_\gamma^2 } { \left( 2 M^2_{H^\pm} + \lambda_1 f^2 \right)^{\! 2} } } \; \, \right]  ~~,
\ee
where we neglected terms suppressed by $v^2/f^2$.
Expanding in $s_\gamma^2$, and using Eq.~(\ref{eq:light_Higgs_mass}), we find 
\be
M_{A_1} \simeq \frac{f}{v} \, M_h    ~~.
\label{MA}
\ee
Even though $A_1$ is much heavier than the Higgs boson, it is substantially lighter than the other composite scalars.

\section{$U(3)_L$ breaking from electroweak interactions}\setcounter{equation}{0}
\label{sec:EW}

In Section 2 we have assumed that the mass and quartic terms in the potential respect the $U(3)_L$ symmetry, and the only explicit $U(3)_L$ breaking comes from tadpole terms. Other explicit $U(3)_L$ breaking effects, such as the $SU(2)_W \times U(1)_Y$ gauge interactions, can feed into the mass and quartic terms through loops. In this section we study the effect 
of these additional $U(3)_L$ breaking effects on $M_h$. 

We parametrize the $U(3)_L$ breaking mass and quartic terms as
\begin{eqnarray}
\Delta V_{\rm breaking} &=& \frac{\kappa_1}{2} \left[(H_t^\dagger H_t)^2+  (H_\chi^\dagger H_\chi)^2+2(H_t^\dagger H_\chi)(H_\chi^\dagger H_t ) \right] + \frac{\kappa_2}{2}  \left( H_t^\dagger H_t+ H_\chi^\dagger H_\chi  \right)^2 \nonumber \\ [1mm]
&& +  \, \kappa'_1 \left[ H_t^\dagger H_t \phi_t^\dagger \phi_t + H_\chi^\dagger H_\chi \phi_\chi^\dagger \phi_\chi +  \left(H_t^\dagger H_\chi \phi_\chi^\dagger \phi_t + {\rm H.c.}  \right) \right] \nonumber \\  [2mm]
&&+ \, \kappa'_2  \left( H_t^\dagger H_t+ H_\chi^\dagger H_\chi \right) \left(\phi_t^\dagger \phi_t + \phi_\chi^\dagger \phi_\chi   \right) \nonumber \\  [2mm]
&& + \, \Delta M_{tt}^2 H_t^\dagger H_t + \Delta M_{\chi \chi}^2 H_{\chi}^\dagger H_\chi +  \left(\Delta M_{\chi t}^2 H_\chi^\dagger H_t 
+ {\rm H.c.}  \right)  ~~,
\label{eq:u3breakingmass}
\end{eqnarray}
where we assumed again, for simplicity, that the quartic terms are $U(2)_R$ symmetric. 
 It is straightforward to repeat the analysis of Section 2 by including Eq.~(\ref{eq:u3breakingmass}). 
 To leading order in $s_\gamma$,  $v^2/f^2$, and $U(3)_L$ breaking, the correction to $M_h^2$ is
\begin{equation}
\Delta M_h^2 \simeq \left( \kappa_{12}  - \frac{5}{2} \kappa'_{12}  - \frac{\Delta M_{\chi\chi}^2}{f^2} \right) v^2 \, ~,
\label{eq:mhcorrections}
\end{equation}
where $\kappa_{12}  \equiv \kappa_1 +  \kappa_2$, $\kappa'_{12}  \equiv \kappa'_1 +  \kappa'_2$.
The corrections from $\Delta M_{tt}^2$ and $\Delta M_{\chi t}^2$ are suppressed
because the Higgs boson resides mostly in $H_{\chi}$.  The correction of Eq.~(\ref{eq:mhcorrections}) 
 can be most easily seen in the limit $s_\gamma \to 0$ and 
 $\Delta M_{\chi t}^2=0$, where $\varphi_t$ and $h_t$ decouple and the mass matrix (\ref{eq:cpeven_mass_matrix}) becomes block-diagonal; we then only need to diagonalize the $2\times 2$ mass matrix:
\be   
\left( \ba{cc}  \displaystyle
\left(\lambda_{12} + \kappa_{12} \right) v^2 & \left(\lambda_{12} + \kappa'_{12} \right) uv \\ [0.3cm]
\left(\lambda_{12} + \kappa'_{12} \right) uv  & 
\ \ -\Delta M_{\chi\chi}^2 + \left(\lambda_{12}  - \frac{1}{2} \kappa'_{12} \right) u^2 - \frac{1}{2} \left(\kappa_{12} -\kappa'_{12} \right) v^2 
\ea\right)  ~~,
\ee
where $\lambda_{12} \equiv \lambda_1+\lambda_2$.
In order to keep the Higgs boson light, and to have the correct vacuum, we need
\be
\left| \kappa_{12}  - \frac{5}{2} \kappa'_{12}  - \frac{\Delta M_{\chi\chi}^2}{f^2}  \right| <  \frac{\lambda_1}{2 \xi^2}  ~~.
\ee 
The contribution due to the mass-squared splitting between $\phi_\chi$ and $H_\chi$ was presented in Refs.~\cite{Chivukula:1998wd,Dobrescu:1999gv}. 

In our model, the additional $U(3)_L$ breaking effects (besides the tadpole terms) come from the $SU(2)_W\times U(1)_Y$ gauge interactions. They contribute to both the mass and quartic terms. 
In the case where the gauge loops are the dominant source of $U(3)_L$ breaking, we have $\Delta M_{tt}^2= \Delta M_{\chi\chi}^2$ and $\Delta M_{\chi t}^2=0$ because the gauge interactions are $U(2)_R$ symmetric.
Only $H_\chi$ and $H_t$ transform under $SU(2)_W\times U(1)_Y$  while $\phi_\chi$ and $\phi_t$ are singlets. The $SU(2)_W\times U(1)_Y$  gauge loops split the masses of  $H_{\chi (t)}$ and $\phi_{\chi (t)}$, analogously to the mass splitting between the charged and neutral pions due to the electromagnetic interaction. This contribution is quadratically divergent  and needs to be cut off. In the case of $\pi^+ - \pi^0$ mass difference, the cutoff is effectively provided by the $\rho$ meson mass~\cite{Das:1967it} from the Weinberg sum rules~\cite{Weinberg:1967kj}. Based on this analogy, we denote the cutoff by $M_{\rho}$, which is set by the mass of some (presumably vector) state in this theory. The 1-loop splitting is then given by
\begin{equation}
\Delta M_{\chi\chi}^2=  \frac{3}{64 \pi^2} \left( 3 g_2^2 + g_1^2\right) M_\rho^2  ~~,
\label{eq:EWsplitting}
\end{equation}
where $g_2$ and $g_1$ are the $SU(2)_W\times U(1)_Y$ gauge couplings.
This mass splitting implies that $M_h^2$ receives a correction (at the chiral symmetry breaking scale $f$) of
\begin{equation}
\Delta M_{h\, ({\rm mass})}^2  \approx  - \left( 0.079\,  v \, \frac{M_\rho}{f} \right)^{\! 2} \, ,
\label{eq:mhEW}
\end{equation}
where the gauge couplings are evaluated at 10 TeV.
This effect reduces the Higgs boson mass and can be quite significant if $M_\rho \gg f$. For example, for $M_\rho = 5 f$ this reduces the effective quartic Higgs coupling, $\lambda_h$, by $0.16$ at the chiral symmetry breaking scale $f$. 

The $SU(2)_W \times U(1)_Y$ gauge interactions also generate the additional quartic interactions involving $H_t$ and $H_\chi$. 
The dominant 1-loop contribution can be estimated to be 
\begin{equation}
\frac{\kappa_{1(2)}}{\lambda_{1(2)}} \simeq 2\, \frac{\kappa'_{1(2)}}{\lambda_{1(2)}} \simeq \frac{ 3 }{16 \pi^2} \left(3 g_2^2 +g_1^2 \right) \ln \left(\frac{M_\rho}{\mu}\right)    ~~,
\label{eq:1loopdlambda}
\end{equation}
where we have assumed the same cutoff as in Eq.~(\ref{eq:EWsplitting}), and the renormalization scale $\mu$ should be taken around the heavy scalar states in the spectrum.  We have also neglected the small $g_{1,2}^4$ contributions. Note that 
Eq.~(\ref{eq:1loopdlambda}) is valid only for $\mu < M_\rho$.
The corrections to the quartic couplings give a correction to the Higgs squared mass:
\begin{eqnarray}
\Delta M_{h\, ({\rm quartic})}^2  &\simeq&  -\frac{3}{64 \pi^2} \, \lambda_1 \left(3 g_2^2 +g_1^2 \right) \, v^2 \ln \left(\frac{M_\rho}{\mu} \right)
 \nonumber \\ [1mm]
&\approx &  -0.16 \, v^2 \,  \frac{\lambda_1}{2\xi^2} \left(\frac{\xi}{3.6}\right)^{\! 2} \ln\left(\frac{M_\rho}{\mu}\right)  \, ~.
\label{equation:mhEWquartic}
\end{eqnarray}
This contribution is also negative, so that the electroweak interactions only reduce the Higgs boson mass. Thus, the Higgs mass formula obtained in the previous section [see Eqs.~(\ref{eq:Higgsmass})-(\ref{eq:mhmax})]
provides an upper bound on the Higgs mass in the absence of other $U(3)$ breaking effects. If the cutoff $M_\rho$ is too large, the effective quartic coupling of the light Higgs can turn negative which implies that we are expanding around the wrong vacuum. 
This puts an upper limit on $M_\rho$, which depends on other parameters, as discussed in the next Section.

\section{Higgs mass and the heavy state spectrum}\setcounter{equation}{0}
\label{sec:spectrum}

In the previous sections we derived the approximate expression for the Higgs mass and the corrections from the electroweak gauge loops. In this section we perform a numerical study of the Higgs mass and show that the exact numerical result agrees well with the analytic approximations. After fixing $M_h =126$ GeV, we calculate the masses of the other composite scalars. 

We start with an enumeration of the parameters of this model.
In Eqs.~(\ref{eq:Vtop}) and (\ref{eq:V}), our model contains the following parameters:
\begin{equation}
\xi,~~ \lambda_1,~~ \lambda_2,~~ M^2_{tt},~~ M^2_{\chi\chi},~~ M^2_{\chi t},~~ C_{\chi t},~~ C_{\chi \chi}. \label{eq:paras-1}
\end{equation}
One of the parameters ($M^2_{\chi t}$ or $C_{\chi\chi}$) is not independent due to the freedom of the $U(2)_R$ rotation. 
After minimizing the potential, the mass and tadpole terms can be written in terms of the VEVs, quartic couplings and  the charged Higgs mass $M_{H^\pm}$ through Eqs.~(\ref{eq:sol_Mxx})-(\ref{eq:charged_Higgs_mass}). The explicit $U(3)_L$ breaking from the electroweak gauge loops discussed in Section~\ref{sec:EW} introduces one more parameter $M_\rho$, the cutoff of the electroweak gauge loop. As a result, the spectrum is fully determined by the following eight parameters:
\begin{equation}
\xi,~~ \lambda_1,~~ \lambda_2,~~ M_{H^\pm},~~ v,~~ f,~~ s_\gamma,~~ M_\rho. \label{eq:paras-4}
\end{equation}
Two of these eight parameters are fixed by the weak scale and the top mass. 
 To produce the correct $m_t$, we use the SM 1-loop RG equations to evolve the top Yukawa coupling $y_t$ to the scale of the heavy fermion mass $m_{t'}$, and use it to solve for $s_\gamma$, which in the lowest order is given by Eq.~(\ref{sgamma}).  The running top Yukawa coupling in the $\overline{\rm MS}$ scheme at the scale $m_t$ corresponds to $m_t (\mu = m_t)\approx 160$~GeV~\cite{Langenfeld:2009wd}.
 For a set of input values of the other six parameters, which are taken to be
\begin{equation}
\xi,~~ \lambda_1/(2\xi^2),~~ \lambda_2/\lambda_1,~~f, ~~M_{H^\pm}/f,~~ M_\rho/f, \label{eq:paras-5}
\end{equation}
one can calculate the masses and couplings at scale $m_{t'}$.  We choose the ratios of couplings or mass scales as the independent parameters because they are more convenient and better constrained.  To calculate $M_h$, we match the theory to the SM at scale $m_{t'}$, compute the quartic Higgs coupling $\lambda_h$, and then evolve $\lambda_h$ down to the weak scale.

To make a generic prediction for $M_h$, we first examine the expected ranges of input parameters listed in Eq.~(\ref{eq:paras-5}). The $U(3)_L$ symmetry breaking scale must satisfy $f>v$. In fact, as we will see later, $f$ is constrained to be much larger than $v$ by the precision measurement of weak-isospin violating effects. On the other hand, larger $f$ also means more fine tuning for the weak scale. We will consider $f$ up to $10$~TeV to avoid excessive fine-tuning. For the effective theory to be a valid description, the states in the theory should have masses below the cutoff scale. Therefore we will take $M_{H^\pm} < 4\pi f \sim \Lambda$. Similarly, the cutoff of  electroweak gauge loops also satisfies  $M_\rho < 4\pi f$. The ranges of the coupling ratios are discussed in the Appendix and they are expected to be $0.4 \lae \lambda_1/(2\xi^2) \lae 1$ and $-0.2 \lae \lambda_2/\lambda_1 \lae 0$. The Yukawa coupling $\xi$ is expected to be $\sim 3-4$ in a strongly coupled theory. The 1-loop estimate from the compositeness condition $Z_\Phi(\Lambda)=0$ gives
\begin{equation}
\xi^2 \simeq \frac{8\pi^2}{\overline{N} \ln ( \Lambda/ m_{t'})}  ~~,
\end{equation}
where $\overline{N} = N_c =3$ if only fermion loops are included, and $\overline{N} = N_c +5/2 =11/2$ if both fermion and scalar loops are included.
For $\Lambda/ m_{t'} =3 \, (10)$,
 $\xi \approx 4.9 \, (3.4)$  for $\overline{N}=3$, and $\xi \approx 3.6 \, (2.5)$ for $\overline{N}=11/2$. In our numerical study, we use $\xi = 2\pi/\sqrt{3} \approx 3.6$~\cite{Elias:1984zh} as the standard reference value. For comparison, in QCD a constituent mass of the up and down quarks $m_p/3 \approx 313$~MeV corresponds to a Yukawa coupling $\sim 3.4$, which is close to our estimate.

\begin{figure}[t!]
\centering \hspace{-0.4cm}
\includegraphics[width=10.6cm]{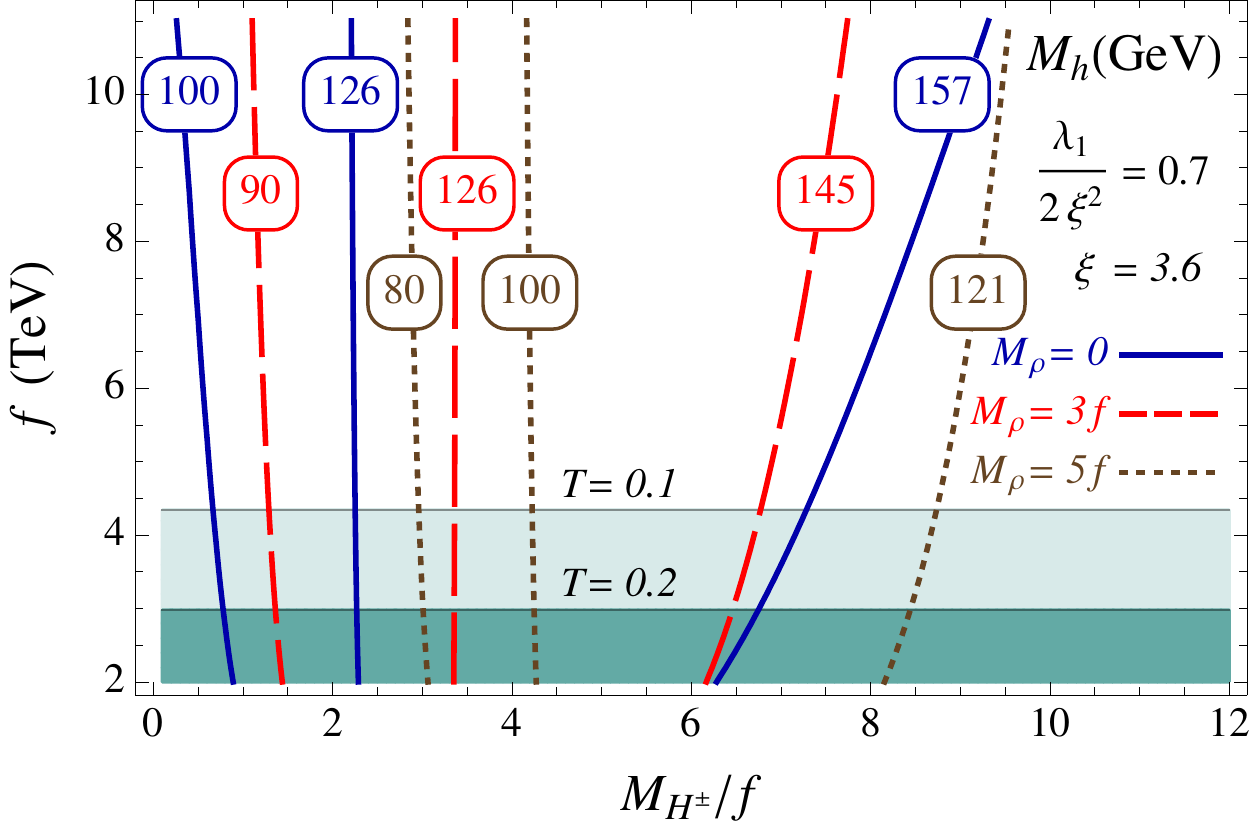}   
\caption{Contours of Higgs boson mass (labelled in GeV), for $\xi=3.6$, $\lambda_1/2\xi^2=0.7$ and $\lambda_2 =0$.    
The solid (blue) lines correspond to no electroweak corrections ($M_\rho =0 $, see Section~\ref{sec:EW}), while the dashed (red) and dotted  (brown) lines correspond to $M_\rho/f = 3 $ and 5, respectively.
The dark and light shaded regions correspond to $T>0.2$ and   $0.1<T<0.2$, respectively, with the $T$-parameter  given by Eq.~(\ref{eq:T}).}
\label{fig:master}
\end{figure}

In Fig.~\ref{fig:master} we show the Higgs boson mass as a function of the dimensionful
parameters $M_{H^\pm}$, $f$ and $M_\rho$ by fixing the dimensionless couplings to some typical values, $\xi=3.6$, $\lambda_1/(2\xi^2)=0.7$ and $\lambda_2/\lambda_1=0$.  We see that $M_h=126$~GeV can be obtained with reasonable parameters of our model. Higgs mass increases as $M_{H^\pm}$ increases, as expected from the approximate analytic formula Eq.~(\ref{eq:light_Higgs_mass}). The dependence on $f$ is mild because it only enters the higher order corrections for fixed $M_{H^{\pm}}/f$ and affects the starting point of RG running of the couplings. On the other hand, $M_h$  decreases as the cutoff $M_\rho$ of the electroweak gauge loops increases due to its negative contribution to $\lambda_h$. For this set of $\{\xi,\, \lambda_1,\, \lambda_2\}$, we see that the correct Higgs boson mass is close to the upper bound for $M_\rho \approx 5 f$ and hence we cannot have $M_\rho > 5f$. This upper bound depends on the values of the coupling parameters, which will be examined later.

In Fig.~\ref{fig:master} we also include contours of the $T$ parameter~\cite{Peskin:1991sw}, 
which measures the weak-isospin violation and constitutes the strongest constraint on this model. The Higgs field arises as the pNGBs of the broken $U(3)_L$ symmetry, which does not contain a custodial $SU(2)$ symmetry. As a result, the dimension-6 operator $|H_\chi^\dagger D_\mu H_\chi|^2$, which represents the weak-isospin violation, is expected based on na\"{i}ve dimensional analysis to have a coefficient $\sim 16\pi^2/\Lambda^2 \sim 1/f^2$.  In our model, the scalar fields are composite degrees of freedom and only exist below the compositeness scale. All low-energy operators containing derivatives of scalars, including the kinetic term and the weak-isospin violating operator $|H_\chi^\dagger D_\mu H_\chi|^2$ operator, are generated by fermion loops~\cite{Hill:1991at,Suzuki:1989si}, 
which can be calculated in the leading $N_c$ approximation. The leading contribution to the $T$ parameter is therefore captured by the loops involving the $t'$ quark~\cite{Dobrescu:1997nm, Chivukula:1998wd, Dobrescu:1999gv},
and is given by \cite{Chivukula:1998wd}
\begin{equation}
T = \frac{3 s^2_L}{16\pi^2 \alpha v^2} \left[ s^2_L m^2_{t'} + 4(1-s^2_L) \frac{m^2_{t'} m^2_t}{m^2_{t'}-m^2_t} \ln{\left(\frac{m_{t'}}{m_t}\right)} - (2-s^2_L)m^2_t \right] ,   \label{eq:T}
\end{equation}
where $s_L$ is the sine of the left-handed mixing angle, given in Eq.~(\ref{eq:thetaL}). 
This is equivalent to calculating the coefficient of the $|H_\chi^\dagger D_\mu H_\chi|^2$ operator, with the Higgs field replaced by its VEV.
Contributions to the $T$ parameter from loops with heavier scalars (which also come from fermion loops in the UV theory, but subleading in $N_c$) are very small compared to the contribution from fermion loops calculated here. The contribution to the $S$ parameter is negligible because we have added only vector-like quarks to the SM.

In the above discussion of the $T$ parameter we have assumed that the strong dynamics responsible for the composite Higgs sector includes only 4-fermion operators
of the NJL type \cite{Dobrescu:1999gv}, which are equivalent to the product of a left-handed current and a right-handed one (LR). If these 
arise from a spontaneously broken gauge symmetry, then 
they are accompanied by LL and RR 4-fermion operators that contribute to the coefficient of the $|H_\chi^\dagger D_\mu H_\chi|^2$ operator \cite{Chivukula:1998uf}.
The ensuing correction to the $T$ parameter, coming from diagrams with two or more 
fermion bubbles, is expected to be somewhat smaller than our result in Eq.~(\ref{eq:T}). In what follows we will ignore these model-dependent corrections related to UV physics.

\begin{figure}[t!]
\centering
\includegraphics[height=6.3cm]{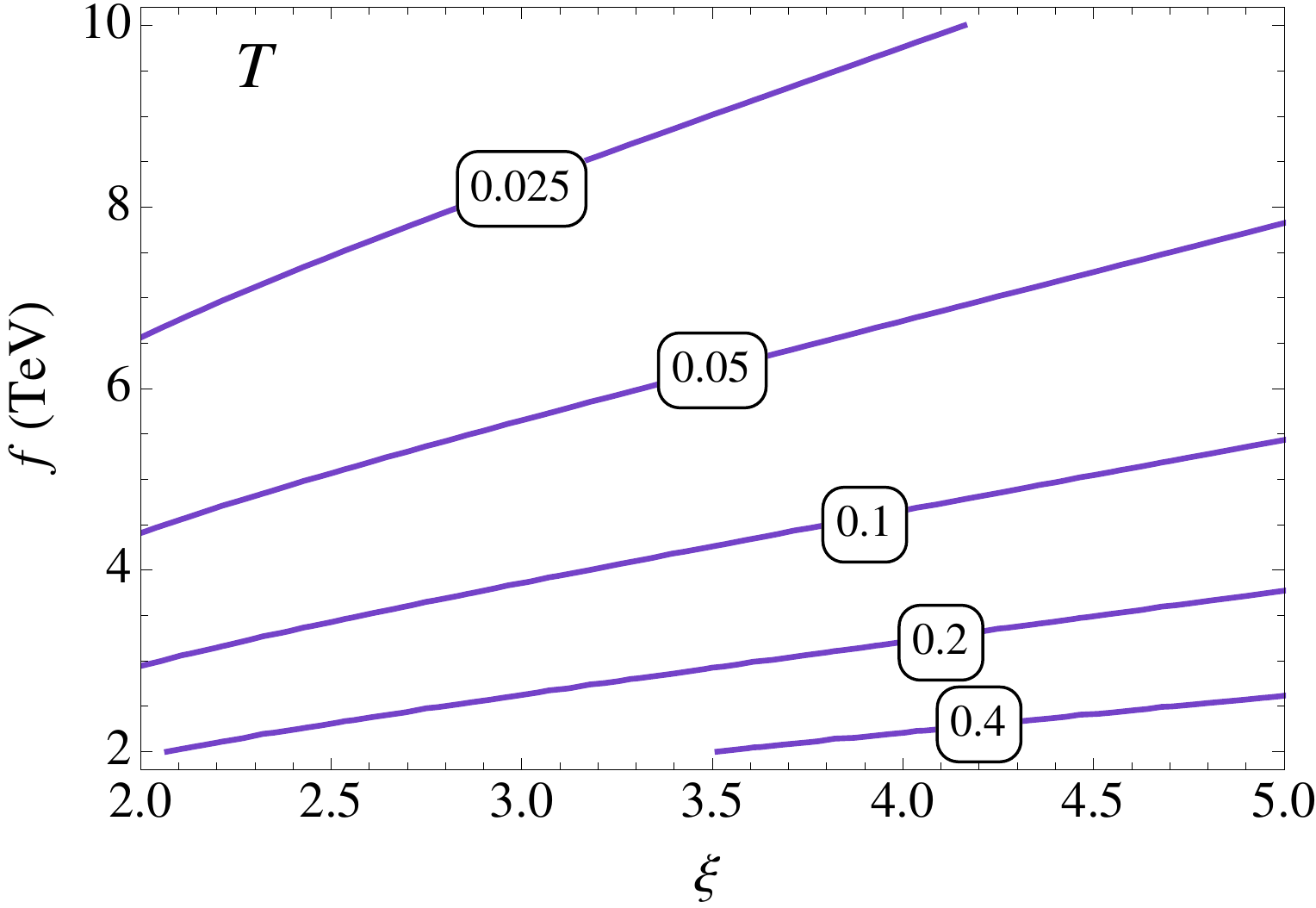}
\caption{Contour plot of the $T$ parameter in the $\xi-f$ plane, given by Eq.~(\ref{eq:T}).  The $T$ parameter is roughly proportional to $1/f^2$. The constraint from the electroweak fit ($T<0.15$ at the 95\% CL) implies $f \gg v$.}
\label{fig:T}
\end{figure}

Eq.~(\ref{eq:T}) only depends on $\xi$, $f$, $v$ and $m_t$, where the last two are fixed by their experimental values.  With $m_{t'}\approx {\xi f}/{\sqrt{2}}$, $s_L \approx {v}/{f}$ [Eqs.~(\ref{eq:mt'}), (\ref{eq:thetaL})], one can see that all three terms in Eq.~(\ref{eq:T}) are roughly proportional to $1/f^2$, while they have different dependence on $\xi$.  Eq.~(\ref{eq:T}) can be rewritten in terms of $\xi$ and $f$ as
\begin{equation}
T \approx \frac{3}{16\pi^2 \alpha f^2} \left[ \frac{v^2 \xi^2}{2} + 4 m^2_t \ln \left(\frac{\xi f}{\sqrt{2} m_t}\right) -2 m^2_t \right]  ~~.  \label{eq:T2}
\end{equation}
By using the low energy values of the Yukawa coupling $\xi$, the running effect from the compositeness scale to the fermion masses is also included. This effect reduces the $T$ parameter somewhat compared to the na\"{i}ve estimated value [$\sim  v^2/(2\alpha f^2)$] and renders a slightly milder constraint.
Contours of $T$ in the $(f,\xi)$ plane are shown in Fig.~\ref{fig:T}.  From the current electroweak fit~\cite{Baak:2012kk}, the $68\%$ and $95\%$ bound roughly correspond to $T=0.1$ and $T=0.15$ (for $S=0$).  For $\xi=3.6$, these bounds translate to $f \gtrsim 4.3$ TeV at $68\%$ CL, and $f \gtrsim 3.5$ TeV at $95\%$ CL.

\begin{figure}[t!]
\centering \hspace{-0.4cm}
\includegraphics[width=11.9cm]{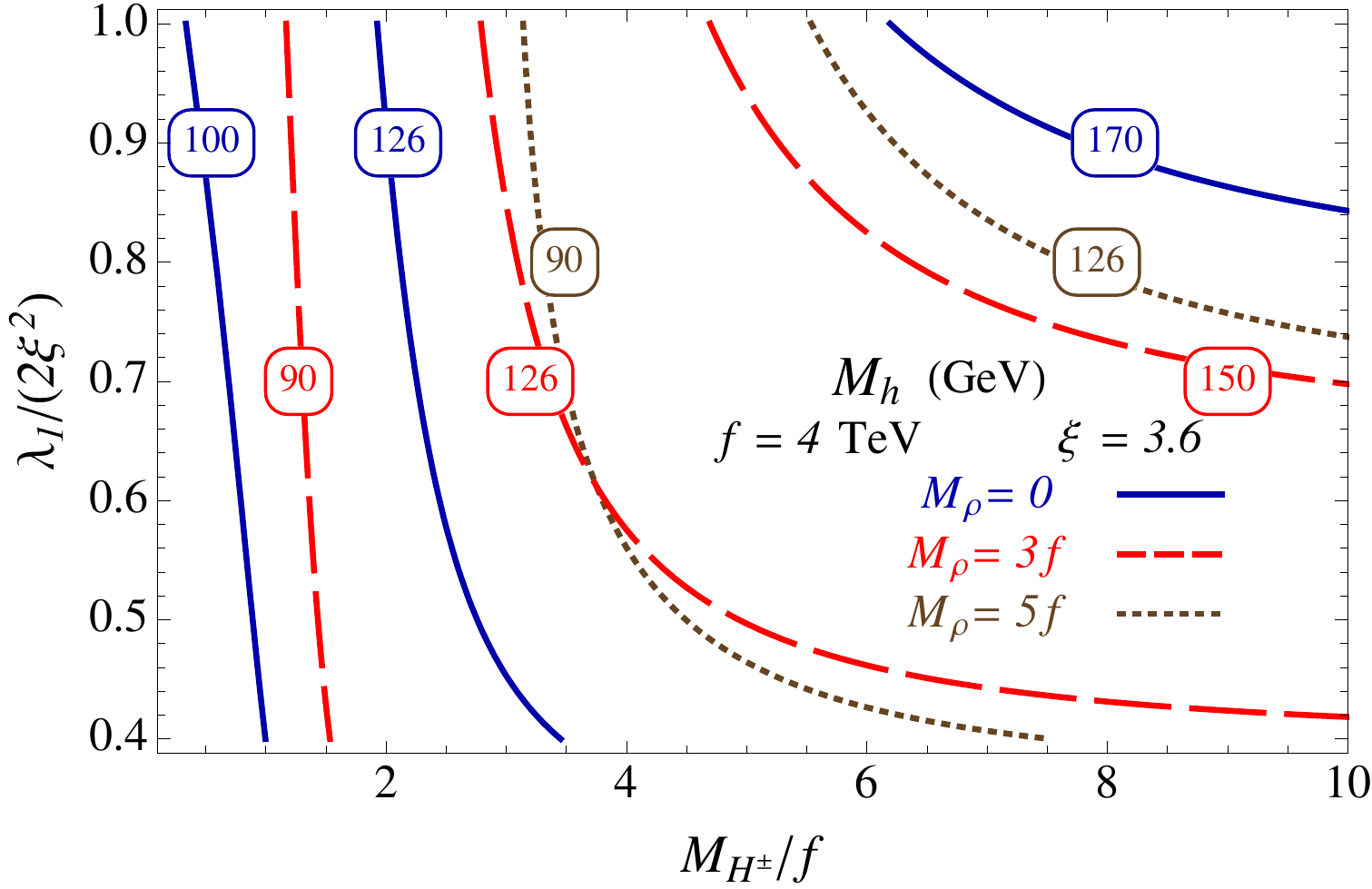}
\caption{Contours of Higgs boson mass (labelled in GeV), for $\xi=3.6$, $\lambda_2 =0$, and $f=4$~TeV.    
The solid (blue) lines correspond to no electroweak corrections ($M_\rho =0 $, see Section~\ref{sec:EW}), while the dashed (red) and dotted  (brown) lines correspond to $M_\rho/f = 3 $ and 5, respectively.}
\label{fig:freemha}
\end{figure}

Among the six parameters of Eq.~(\ref{eq:paras-5}), the Higgs boson mass in the leading order is only sensitive to $\lambda_1/(2\xi^2)$, $M_{H^\pm}/f$ and $M_\rho/f$.  The dependence on the other parameters are expected to be mild. To study the $M_h$ dependence on the more sensitive parameters, we fix $\xi=3.6$, $\lambda_2=0$, and $f=4$~TeV (which corresponds to $T$=0.12, close to its lower bound), and make the contour plot of the Higgs boson mass in the 
$\lambda_1/(2\xi^2)$-versus-$M_{H^\pm}/f$ plane for several different values of $M_\rho/f$ (Fig~\ref{fig:freemha}).
As expected, a larger $M_h$ occurs for larger $M_{H^\pm}/f$, $\lambda_1/(2\xi^2)$ and smaller $M_\rho/f$. There is an upper limit $M_h \lesssim 175$~GeV even for extreme values of these parameters. This is close to our estimate from the analytic formula, Eq.~(\ref{eq:mhmax}). 

A lower bound on $M_h$ follows from the condition that the quartic coupling $\lambda_h$ is positive at the matching scale $m_{t'}$. 
Otherwise of our vacuum is not a minimum of the tree-level potential, and  the universe is more likely to end up in wrong vacuum.
Imposing the boundary condition $\lambda_h=0$ at the scale $m_{t'}\simeq \xi f/\sqrt{2}$, and using the SM 1-loop RG equations to evolve
 $\lambda_h$ down to the weak scale, we find that the physical $M_h$ grows monotonically from 80 GeV for
 $m_{t'} = 6$ TeV (corresponding, {\it e.g.}, to $f = 3.5$ TeV, $\xi = 2.5$)
 to 90 GeV for  $m_{t'} = 25$ TeV.  
Thus, the lower bound on $M_h$ is around 80 GeV. 
We conclude that, in our composite Higgs model, the Higgs boson mass is constrained to be in the 
\be
80 \; {\rm GeV} < M_h < 175  \; {\rm GeV} 
\ee
range, with the upper limit significantly tighter for most of the parameter space (as shown in Figs.~\ref{fig:master} and \ref{fig:freemha}); 
the measured $126$~GeV Higgs mass sits comfortably in the middle of this range.

So far we have considered the Higgs boson mass dependences on the more sensitive parameters. To check how the Higgs mass varies with the less sensitive parameters, we show in Fig.~\ref{fig:freemhb} 
the Higgs mass as a function of $M_{H^\pm}/f$ for several different sets of $\{ \xi,\, f,\, \lambda_2\}$, by fixing $\lambda_1/(2\xi^2)=0.7$ and $M_\rho=3f$. We see that indeed the dependences on these parameters are rather mild. Among them, the Yukawa coupling $\xi$, which enters both the higher order correction in Eq.~(\ref{eq:light_Higgs_mass}) (through $m_{t'}$) and the electroweak gauge loop correction in Eq.~(\ref{equation:mhEWquartic}) [for a fixed value of $\lambda_1/(2\xi^2)$], has a slightly larger effect. The $\lambda_2$ dependence is almost negligible.

\begin{figure}[t!]
\centering
\includegraphics[height=8.1cm]{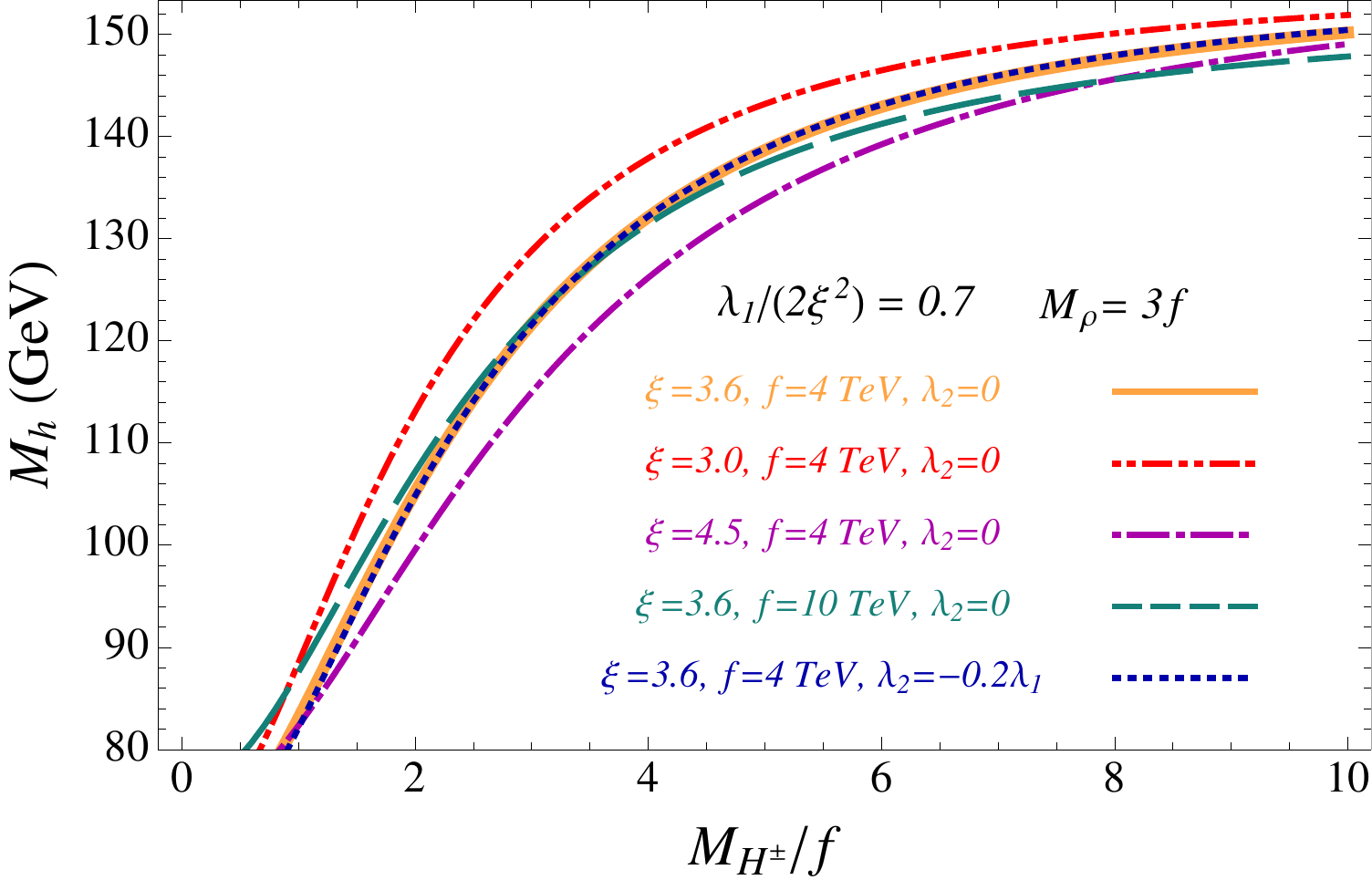}  
\caption{Higgs boson mass as a function of $M_{H^\pm}/f$ for various values of $\xi,\, f,\, \lambda_2$, when $\lambda_1/2\xi^2=0.7$ and $M_\rho /f=3$. 
The different curves are obtained by varying one parameter at a time with respect to the solid (orange) curve. }
\label{fig:freemhb}
\end{figure}

In addition to $M_h$, we are also interested in other predictions of this model, such as the spectrum of heavy states.
This model contains two doublet and two singlet scalars in the effective theory. After electroweak symmetry breaking, apart from the eaten Nambu-Goldstone modes and the SM-like Higgs boson, there are 3 neutral CP-even scalars (denoted by $H_{1,2,3}$ according to the ascending order of their masses), 3 neutral CP-odd scalars (denoted by $A_{1,2,3}$), and a complex charged Higgs boson ($H^\pm$). One CP-odd scalar ($A_1$) is lighter than other heavy states because it is also the pNGB of the broken $U(3)_L $ symmetry and hence its mass is also suppressed by $s_\gamma$. However, its mass is controlled by $f$ rather than $v$, so it is still quite heavy compared to the Higgs boson. The masses of the states coming from $\Phi_t$ receive contribution from both $M_{tt}^2$ and $\lambda_1 f^2$. In the limit of $M_{tt}^2 \gg \lambda_1 f^2$, all these states are expected to be around $M_{H^\pm}$ and decouple from low energy physics.

\begin{figure}[t!]
\centering
\includegraphics[height=7.2cm]{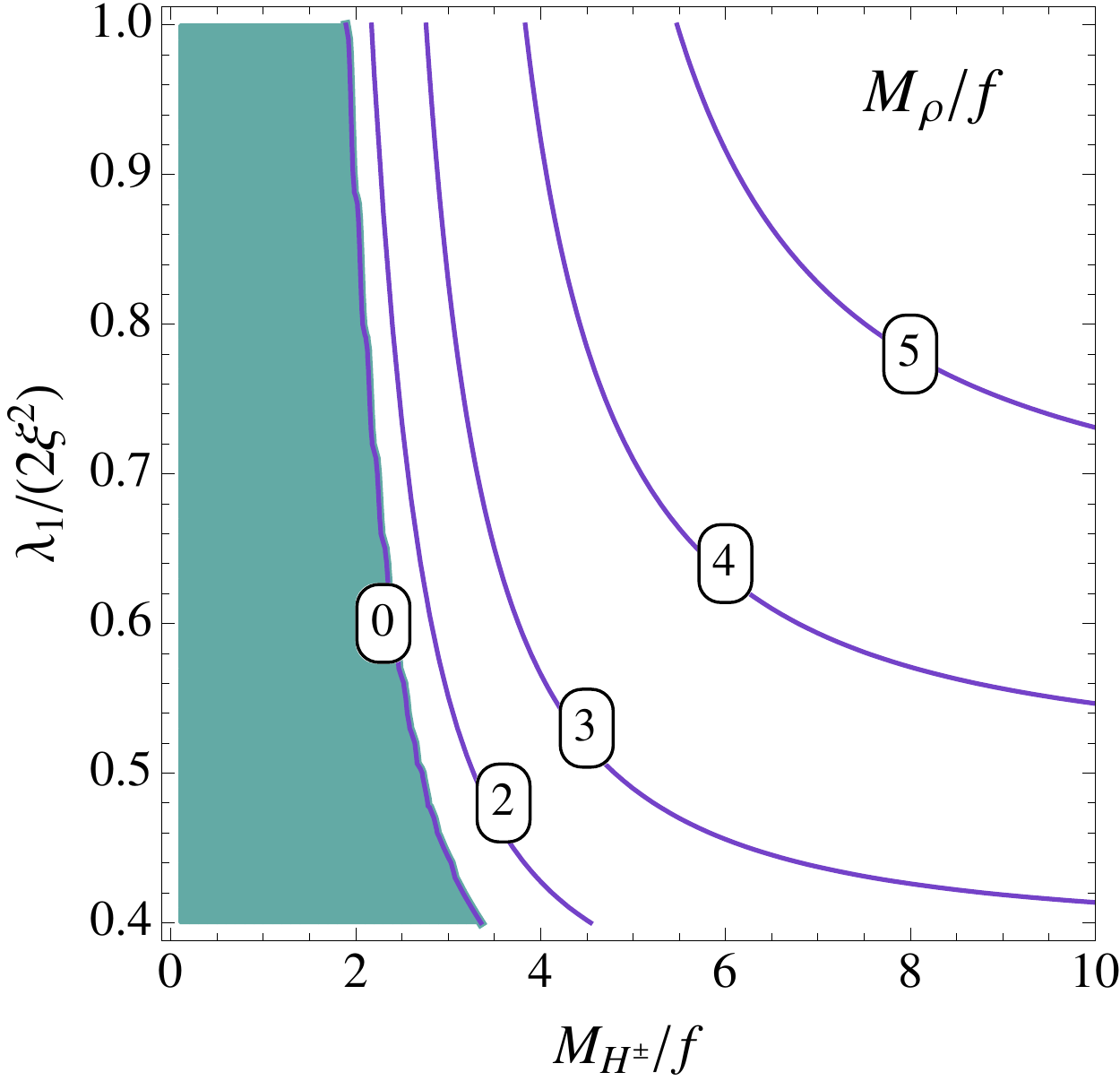}
\includegraphics[height=7.2cm]{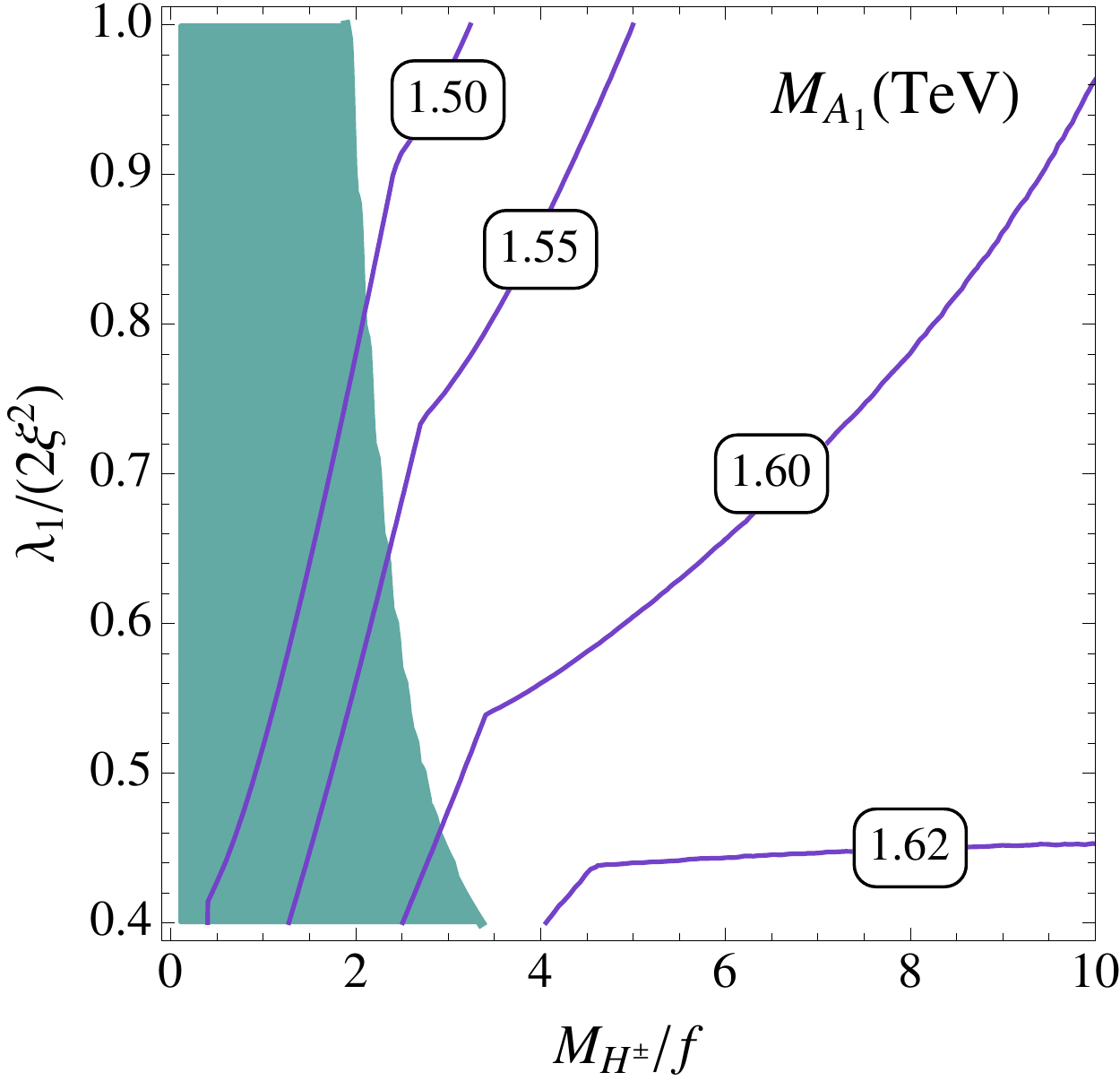}  \\ \vspace{0.3cm}
\includegraphics[height=7.2cm]{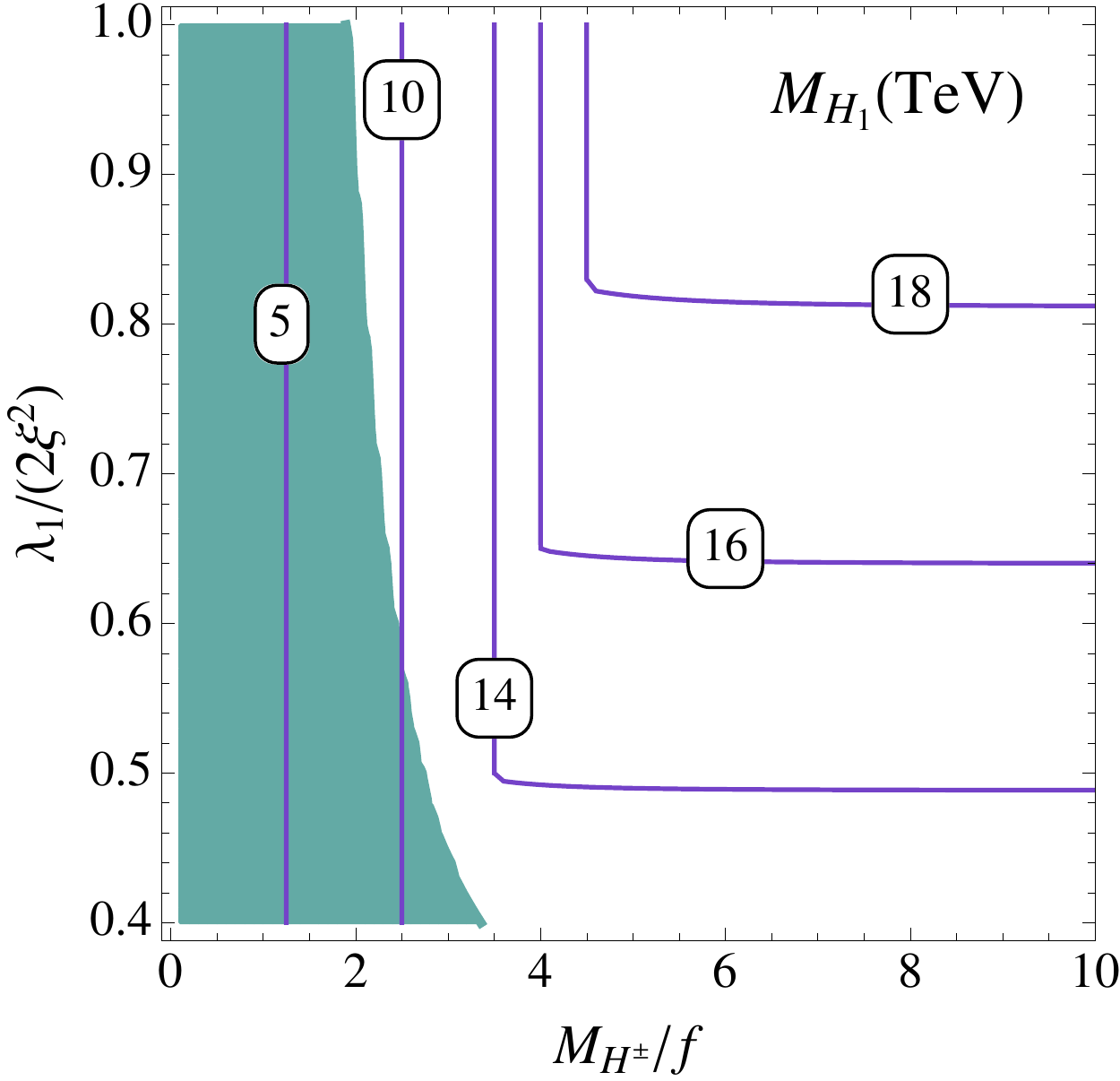}
\includegraphics[height=7.2cm]{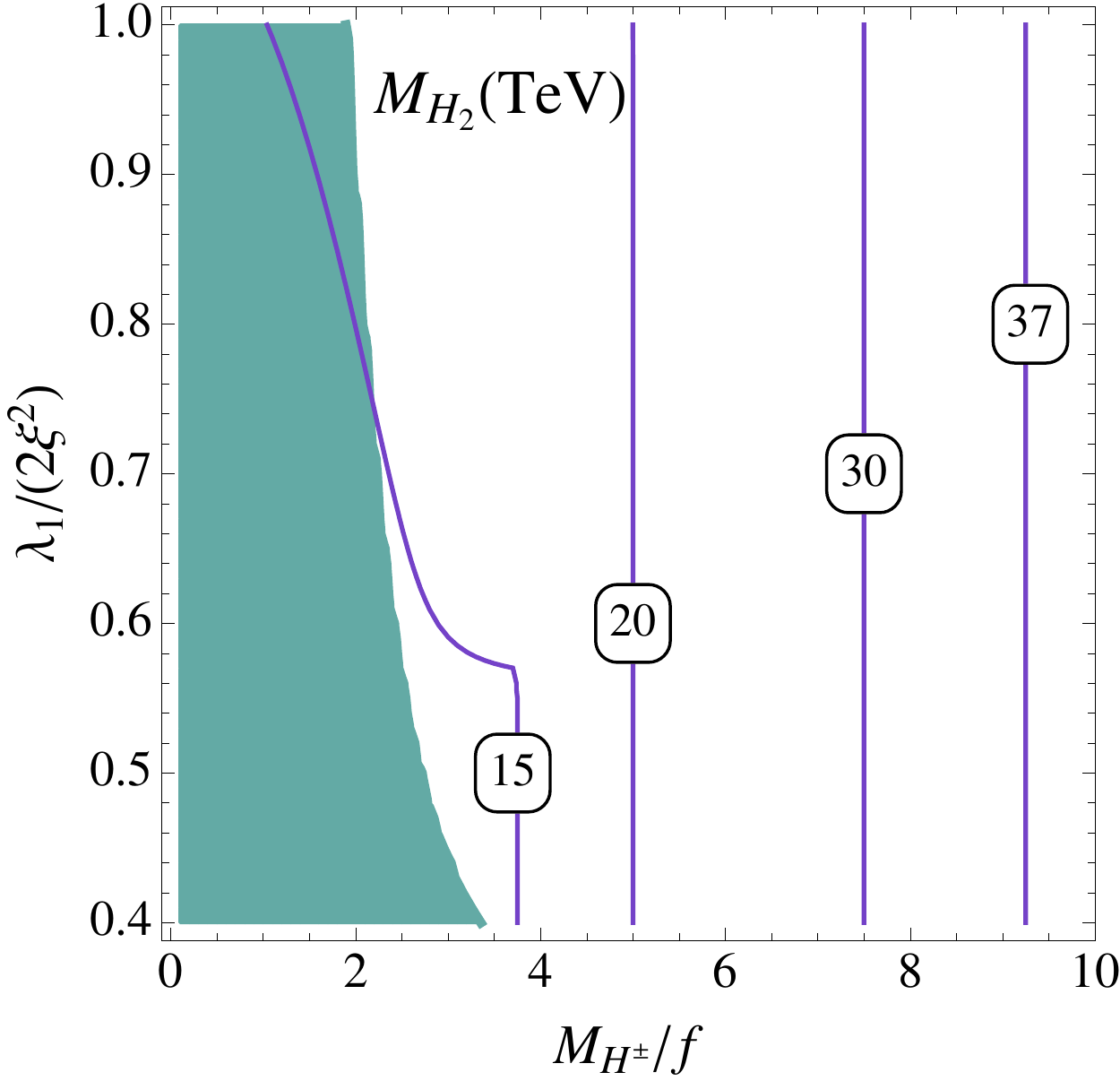}
\caption{Contour plots for $M_\rho/f$, the mass of the lightest CP-odd neutral scalar ($M_{A_1}$) and the masses of the second and third lightest CP-even neutral scalars ($M_{H_1}$ and $M_{H_2}$).  We have fixed $f=4$~TeV, $\xi=3.6$ and $\lambda_2 =0$, which corresponds to $m_{t'}=10.2$~TeV and $T=0.12$. The shaded regions are not consistent with $M_h = 126$ GeV (unless there is additional  $U(3)_L$ breaking with effects opposite to the electroweak corrections).}
\label{fig:fixedmh}
\end{figure}

We use the mass of the discovered Higgs boson to fix one more parameter. We choose it to be $M_\rho$ and plot its required values to obtain the correct Higgs boson mass in the $M_{H^\pm}/f - \lambda_1/(2\xi^2)$ plane in the first panel of Fig.~\ref{fig:fixedmh}. 
The contour $M_\rho/f=0$ represents the case where the explicit $U(3)_L$ breaking from the electroweak gauge loops is absent. $M_h = 126$ GeV cannot be obtained in the region to the left of that contour, unless there is additional explicit $U(3)_L$ breaking from the cutoff scale that has positive contributions to $\lambda_h$.
For small values of $M_{H^\pm}$ or $\lambda_1/(2\xi^2)$, $M_\rho/f$ also needs to be small, which means that additional states responsible for cutting off the electroweak gauge loop contribution are probably required to appear near or even below the scales of our heavy scalars. 

Assuming that $M_\rho$ takes the value which produces $M_h \approx 126$~GeV, we can calculate the masses of the heavy scalars in this theory. We show the masses of the lightest CP-odd scalar ($A_1$) and the first two heavy CP-even neutral scalars ($H_1$, $H_2$) in the other three panels of Fig.~\ref{fig:fixedmh}, for 
the same parameters as in Fig.~\ref{fig:freemha}.
For fixed $M_{H^\pm}/f$, the heavy states scale proportionally to $f$. Since $f=4$~TeV is close to the lower bound allowed by the $T$ parameter, the masses in these plots are close to their lower limits in this model. We see that $A_1$ is indeed much lighter compared to other heavy states, though it is still above 1 TeV. For small values of $M_{H^\pm}/f$, $H_1$ mostly comes from $\Phi_t$ and its mass is close to $M_{H^\pm}$, so the contours run vertically. On the other hand for large $M_{H^\pm}/f$, $H_1$ dominantly consists of $\varphi_\chi$ and its mass is proportional to $\lambda_1$, and is independent of $M_{H^\pm}$. The second pseudoscalar $A_2$ is closely degenerate with $H^\pm$, and the masses of $H_3$, $A_3$ are a little bit higher but are also close. 

\section{Phenomenology} \setcounter{equation}{0}

Within this composite Higgs model, the lightest particle beyond the SM is the CP-odd scalar $A_1$. Its mass is 
larger than $M_h$ by a factor of $f/v$, as shown in Eq.~(\ref{MA}). Taking into account the scale dependence of $M_h$,
the $A_1$ mass $M_{A_1}$ is slightly reduced (see second panel of Fig.~\ref{fig:fixedmh}). The smallest value of $f$ allowed by the electroweak data,
$f = 3.5$ TeV, gives $M_{A_1} = 1.3$ TeV. 
$A_1$ may be singly produced at hadron colliders through gluon fusion. Given that $A_1$ is mostly part of the $\phi_\chi$ singlet,
its coupling to the top quark is suppressed. The dominant contribution to gluon fusion is a $t'$ loop, relying on the $(\xi/\sqrt{2})\,   A_1 \bar{t}' \gamma_5 t'$ coupling.
For $f = 3.5$ TeV, the $A_1$ production cross section  is 
$\sim 0.6 $ fb at the 14 TeV LHC. We obtained this result using FeynRules \cite{Alloul:2013bka} and MadGraph 5 \cite{Alwall:2011uj}
and then multiplying by a factor of 2 to account for large higher order effects.
$A_1$ may decay into $t\bar{t}$ or $h Z$, but in either case the cross section for $pp \to A_1 X$ is too small,
even for a luminosity of 3000 fb$^{-1}$ \cite{ATLAS:2013-003}. 

The other composite scalars as well as the vector-like quark are too heavy to be probed at the LHC. 
A hadron collider at a center-of-mass energy of $O(100)$ TeV would be needed in order to produce them. 
For a typical $m_{t'} = 9$ TeV,
the $t'\bar{t'}$ production cross section at a 100 TeV $pp$ collider (we refer to it as the VLHC) 
is 0.12 fb, based on MadGraph 5, with a 50\% increase to account for higher order effects. 
The VLHC cross section can be as large as  1.8 fb, for  $m_{t'} = 6$ TeV. 
 Besides the usual $t' \to Wb, t h, tZ$ decays \cite{Han:2003wu}, there is an interesting $t' \to t A_1$ decay, followed by $A_1 \to t\bar{t}$ or $h Z$.
Given that the backgrounds relevant at an invariant mass $\gtrsim 6$ TeV are small, it is possible to discover the $t'$ quark at the VLHC even 
with the initial luminosity of $2\times 10^{34}$ cm$^{-2}$s$^{-1}$ \cite{Bhat:2013epd}.

Single $A_1$ production at  the VLHC through gluon fusion would have a cross section around 100 fb, for  $M_{A_1} = 1.3$ TeV. 
Detailed studies of the backgrounds are necessary before deciding whether single $A_1$ production leads to promising channels.

The heaviness of the particles beyond the SM puts this model close to the decoupling limit, so the Higgs couplings are close to the SM predictions. The main correction comes from mixing with the singlet scalar fields, which can be seen in Eq.~(\ref{eq:cpeven_mass_matrix}). We have verified numerically that throughout the allowed parameter space, the couplings of the Higgs boson to SM particles are given, to a very good approximation, by their SM values times  
 the factor $\cos (v/f) \simeq 1- v^2/(2 f^2)$, which is the fraction of the doublet component in the Higgs boson. The deviation from the SM couplings is only 0.2\% for $f =4$~TeV and is inversely proportional to $f^2$. The modifications to the branching fractions are even smaller because the dominant correction is universal. Such small deviations can not be tested at the LHC and are probably even beyond the reach of a future $e^+e^-$ collider. 
Nevertheless, a precise determination of the custodial $SU(2)$ breaking $T$ parameter can help to probe or constrain this model further.

Since the SM-like Higgs field is composite and made of the top quark and the new quark $\chi$, the light SM fermion masses presumably come from some 4-fermion interactions generated above the cutoff scale. Constraints on flavor changing neutral currents (FCNC) limit the coefficients and patterns of these 4-fermion interactions  \cite{Dobrescu:1999gv}. Below the cutoff, our model has two Higgs doublets. General couplings of SM fermions to the two Higgs doublets lead to tree-level FCNCs. However, it is reasonable to assume that the couplings of the SM fermions to heavy Higgs states have a hierarchical structure correlated with the corresponding Yukawa couplings of the SM Higgs boson, due to some approximate flavor symmetries~\cite{Cheng:1987rs}. In that case, the FCNC constraints require the scalars from the other Higgs doublet to be heavier than a few hundred GeV to $\sim 1$ TeV, with the strongest bounds coming from the neutral meson-anti-meson mixings and $\mu \to e\gamma$. (See Ref.~\cite{Atwood:1996vj} for a comprehensive study and review.) With the heavy states around 10 TeV in our model, these FCNC constraints are easily satisfied. 
Only $A_1$ is relatively lighter, but it is mostly part of the $\phi_\chi$ singlet, so that it does not induce any significant FCNC effects.

\section{Conclusions} \setcounter{equation}{0}
\label{sec:conclusions}

We have studied a composite Higgs model based on non-confining dynamics, in which the newly discovered Higgs boson 
is a bound state of a vector-like quark and the left-handed top quark. 
The strongly coupled 4-quark interactions that describe the non-confining dynamics at the compositeness scale $\Lambda$
produce scalar bound states which consist of two $SU(2)_W$-doublets and two gauge singlets \cite{Dobrescu:1999gv}. 
We have shown that if the underlying strong dynamics preserves an approximate $U(3)_L$ chiral symmetry, a SM-like
Higgs doublet arises naturally as the pNGB of $U(3)_L \to U(2)_L$ breaking. 

Explicit  $U(3)_L$ breaking terms produce the correct $m_t$ through the top-seesaw mechanism.
They also give the mass to the SM-like Higgs boson. As a result, the Higgs and top masses are tightly correlated, and satisfy $M_h \lesssim m_t$.
Electroweak effects further reduce $M_h$, so that it is easily compatible with the measured Higgs mass within the natural range of parameter space.

The strongest constraint on this model comes from weak-isospin violation due to heavy quark loops. Requiring $T \lesssim 0.15$ pushes the $U(3)_L$ symmetry breaking scale $f$ above 3.5 TeV, so that some fine-tuning is needed to obtain the weak scale $v \approx 246$ GeV $\ll f$. It also means that most of the new states beyond the SM (except the lightest CP-odd scalar) will have masses around or above 10 TeV, beyond the reach of the LHC. The corrections to the SM Higgs couplings are tiny as the new sector is close to the decoupling limit. Nevertheless, the fact that no new particles or any deviation from the SM has been discovered at the LHC so far suggests that  the SM Higgs sector is somewhat tuned, and the scale of new physics may be higher than previously thought. Our model is certainly consistent with the current experimental observations. It would require a collider beyond the LHC with a center-of-mass energy $O(100)$~TeV to probe the heavy states in this model directly. 

On the other hand, 
 the scale of $U(3)_L$ chiral symmetry breaking may be significantly lowered if the contribution to $T$ from the heavy quark can be cancelled by some additional contribution. This would make the model less tuned and the new states lighter and more accessible. If the model can be extended to include an approximate custodial $SU(2)$ symmetry, for example, by adding a new vector-like quark to mix with the bottom quark, then $f$ can be lowered to $\lesssim 1$~TeV, and the new quarks 
could have masses below 2 TeV and be within the reach of the 14 TeV LHC. 

\bigskip\bigskip

{\bf Acknowledgements}:
We would like to thank Sekhar Chivukula, Markus Luty and John Terning for useful discussions. H.-C.~Cheng and J.~Gu are supported by the Department of Energy (DOE) under contract no.\ DE-FG02-91ER40674. 
Fermilab is operated by the Fermi Research Alliance under Contract No. De-AC02-07CH11359 with the DOE. 
H.-C.~Cheng would like to acknowledge the hospitality of National Center for Theoretical Sciences (North) in Taiwan where part of this work was done.

\appendix
\section{Renormalization group running of couplings}\setcounter{equation}{0}
\label{sec:rge}

In Sec.~\ref{sec:MH} the light Higgs boson mass is shown to be proportional to the ratio of the couplings, $\sqrt{\lambda_1/ (2 \xi^2)}$, in the absence of  $SU(2)_W \times U(1)_Y$ gauge interactions. In the fermion-loop (bubble) approximation, this ratio is predicted to be 1, which corresponds to the well-known result of $m_{\phi}=2 m_f$ in the NJL type model~\cite{Nambu:1961tp}, where $m_\phi$ is the mass of the composite scalar and $m_f$ is the constituent fermion mass after chiral symmetry breaking. The fermion-loop approximation neglects the gauge loop corrections and the back reaction of the scalar self-interactions. It can be viewed as the leading $N_c$ result if the gauge interactions are ignored. The presence of the other interactions will modify this ratio. If the chiral symmetry breaking scale $f$ is tuned to be much smaller than the compositeness scale $\Lambda$, this ratio can also be well-determined due to the infrared fixed point structure of the RG equations~\cite{Bardeen:1989ds}. Quasi-infrared fixed points have been used to predict top quark and Higgs boson masses in some theoretical models with large couplings at high scales
\cite{Pendleton:1980as,ArkaniHamed:2000hv}.
For $f$ not much smaller than $\Lambda$ as in the case we are interested, one cannot trust the RG analysis because the couplings are strong and the logarithms are only ${\cal O}(1)$. Nevertheless, finding the infrared fixed point of the RG equations may still provide us some ideas of the possible range of the relevant coupling ratio $\lambda_1/ (2 \xi^2)$. In the far infrared the couplings become perturbative, and the fixed point can be determined by 1-loop RG equations presented below.

To be general and to identify the fermion and scalar loops, we write down the coupled RG equations of the couplings $\xi$, $\lambda_1$, $\lambda_2$, and QCD strong coupling $g_3$ for an $U(N_L)_L \times U(N_R)_R$ theory:
\begin{eqnarray}
16\pi^2 \frac{d g_3}{d t} &=& - \left(11-\frac{2}{3}N_f \right) g^3_3,  \label{eq:rgg3}  \\
16\pi^2 \frac{d\xi}{dt} &=& \left(\frac{N_L+N_R}{2}+ N_c\right) \xi^3- 3\, \frac{N_c^2 -1}{N_c} g^2_3 \, \xi,   \label{eq:rgxi}     \\
16\pi^2 \frac{d \lambda_1}{dt} &=& 2(N_L+N_R) \lambda^2_1 + 4\lambda_1\lambda_2 + 4 N_c (\xi^2\lambda_1 - \xi^4),  \label{eq:rglam1}  \\
16\pi^2 \frac{d \lambda_2}{dt} &=& 4\lambda^2_1 + 4(N_L+N_R) \lambda_1\lambda_2 + 2 N_L N_R \lambda^2_2 + 4 N_c\xi^2\lambda_2, \label{eq:rglam2}
\end{eqnarray}
where we have ignored the electroweak couplings $g_1$, $g_2$, and the light fermion Yukawa couplings. $N_f$ is the number of quark flavors. These equations can be inferred from Ref.~\cite{Cheng:1973nv}.

These RG equations are in the mass-independent scheme. 
Near the composite scale where the composite scalars dissolve, the scalar masses are large and the scalar loops should decouple \cite{Bando:1990wh}. This justifies the fermion-loop approximation near the compositeness scale if the gauge couplings are relatively small. If we drop the scalar loop contributions ({\it i.e.}, terms without the $N_c$ factor) and ignore the gauge couplings, we obtain
\begin{equation}
16 \pi^2 \frac{d}{dt} \ln \left(\frac{\lambda_1}{\xi^2}\right) = 2 N_c \xi^2 \left(1- \frac{2 \xi^2}{\lambda_1}\right)\, . \label{eq:leadingnc}
\end{equation}
We see that the infrared fixed point corresponds to $\lambda_1 = 2 \xi^2$, agreeing with the result of the fermion-loop approximation. On the other hand, $\lambda_2$ is not generated by the fermion loops. Note that the fermion-loop approximation sums fermion loops to all orders so it applies even to large couplings. As a result, one may treat $\lambda_1 = 2 \xi^2, \, \lambda_2=0$ as the initial condition when the scalar loops become relevant.

It is instructive to derive the approximate IR fixed point analytically for the ratios of couplings. For simplicity we first neglect the QCD coupling $g_3$ because it is much smaller than the other couplings near the cutoff scale. 
We obtain the RG equations for $r \equiv \lambda_2/ \lambda_1$ and $s \equiv \lambda_1/ \xi^2$ by combining Eqs.~(\ref{eq:rgxi})-(\ref{eq:rglam2}):
\begin{eqnarray}
16 \pi^2 \frac{ d \ln r}{dt} &=& 2\lambda_1 \left[  \frac{2}{r} + N_L+N_R + (N_L N_R -2) \, r+\frac{2 N_c}{s^2} \right]\, , \label{eq:rgr} \\ [1mm]
16 \pi^2 \frac{ d \ln s}{dt} &=& \xi^2 \left[2(N_L+N_R +2 r ) \, s   + 2 N_c -N_L -N_R - \frac{4 N_c}{s} \right] \, . \label{eq:rqs} 
\end{eqnarray}
The infrared fixed point is reached when the right-hand side of the equations vanishes:
\begin{eqnarray}
&& \frac{2}{r} + N_L+N_R + (N_L N_R -2) \, r+\frac{2 N_c}{s^2} =0 \, , \label{eq:IR_fixed_1}  \nonumber  \\
&& 2(N_L+N_R + 2 r ) \, s   + 2 N_c -N_L -N_R - \frac{4 N_c}{s} =0 \, . \label{eq:IR_fixed_2} 
\end{eqnarray}
There are multiple solutions to these polynomial equations. The actual IR fixed point has $|r| \ll 1$, so we can further simplify the equations by ignoring the terms proportional to positive powers of $r$, then Eq.~(\ref{eq:IR_fixed_2}) gives
\begin{equation}
s_* \simeq \frac{1}{4}\left( 1-x  + \sqrt{ 1+ 14 x + x^2} \right)  \, ~,
\label{eq:IR_s}
\end{equation}
where $x \equiv 2 N_c/(N_L +N_R)$.
We have chosen the positive solution because both $\xi^2$ and $\lambda_1$ stay positive. Substituting it into Eq.~(\ref{eq:IR_fixed_1}), we obtain
\begin{equation}
r_* \simeq - \frac{x}{N_c \left(1 + x /s_*^2 \right) } \, .
\label{eq:IR_r}
\end{equation}
For $N_L=3,\, N_R=2$ and $N_c=3$, we have 
\begin{equation}
s_* \approx 1 \quad \mbox{and} \quad r_* \approx - 0.2\, ~ .
\label{eq:IRvalues}
\end{equation}

\begin{figure}
\centering
\includegraphics[width=12cm]{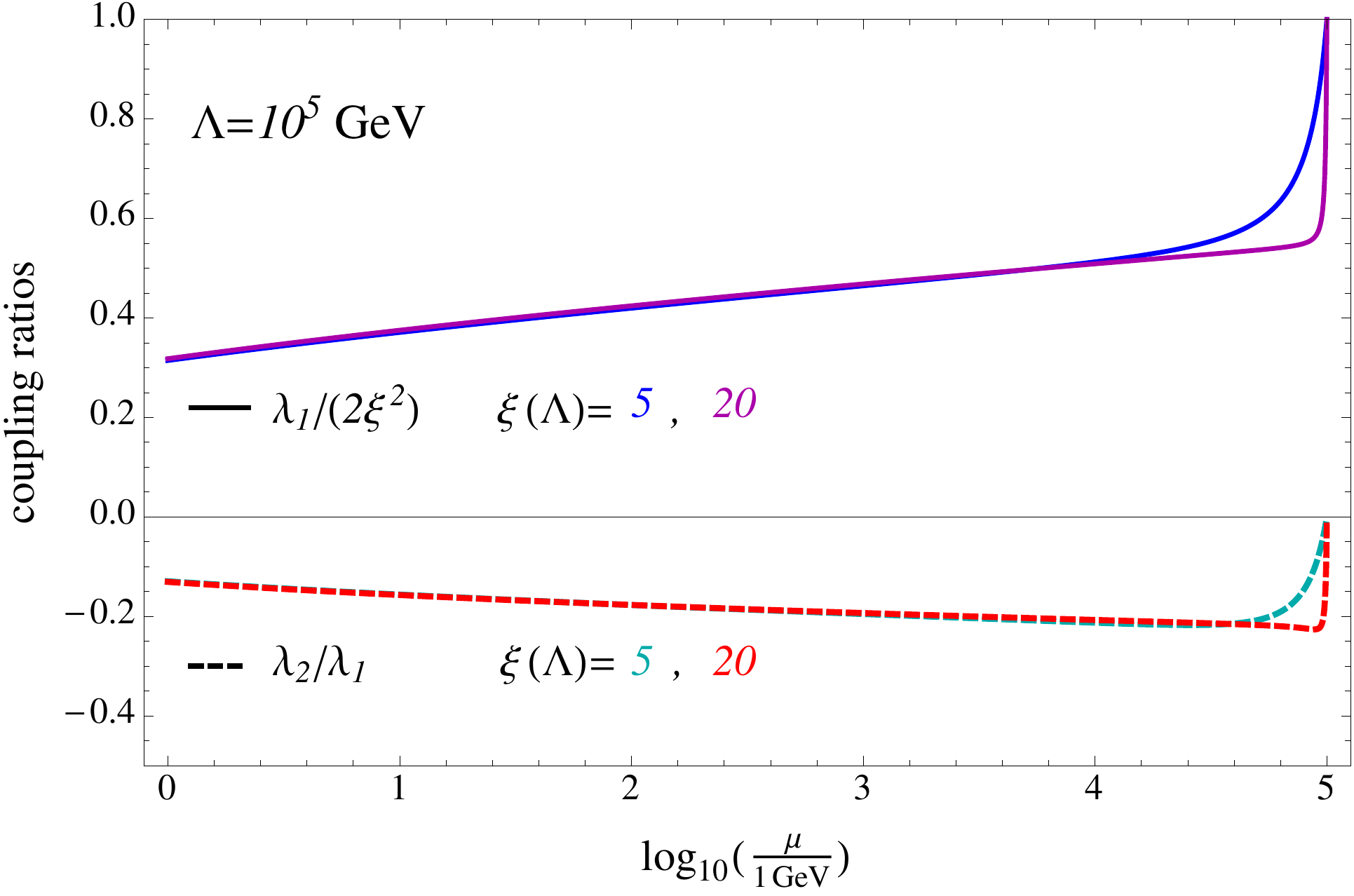}
\caption{One-loop RG evolutions of the coupling ratios $\lambda_1/(2\xi^2)$ and $\lambda_2/\lambda_1$ for initial values  
$\lambda_1/(2\xi^2)=1$, $\lambda_2/\lambda_1=0$ and $\xi=5$ or 20. The horizontal axis is the logarithm of the energy scale.}
\label{fig:rg}
\end{figure}

To check the accuracy of the analytical approximation of the IR fixed point solution, we solve the 1-loop RG equations (\ref{eq:rgg3})--(\ref{eq:rglam2}) numerically. We set the initial condition $\lambda_1 = 2 \xi^2, \, \lambda_2=0$ and choose several different initial values for $\xi$. The results of 1-loop RG running are shown in Fig.~\ref{fig:rg}. We see that the ratios of couplings are quickly driven to the approximate fixed point values given by Eq.~(\ref{eq:IRvalues}), though we should not trust the exact evolution in the beginning due to potentially large higher loop contributions. The infrared value of $r$ is a bit smaller than the approximate result in Eq.~(\ref{eq:IRvalues}) due to the gauge loop contribution from $g_3$.

If the chiral symmetry breaking scale is not far below the compositeness scale, we can not trust the 1-loop RG results. However, if we assume a smooth evolution, the ratios of couplings are expected to lie in between their initial values and the infrared fixed point values:
\begin{equation}
0.4 \lesssim \frac{\lambda_1}{2 \xi^2} \lesssim 1, \quad -0.2 \lesssim \frac{\lambda_2}{\lambda_1} \lesssim 0 \, .
\label{eq:ratio_range}
\end{equation}
We use these ranges in Sections 2 and 4.

\end{document}